\documentclass[superscriptaddress,secnumarabic,
amssymb,amsmath,nobibnotes,aps,prd,showkeys,showpacs,nofootinbib]{revtex4}%
\usepackage{graphicx}
\usepackage{epsf}
\usepackage{bm}
\usepackage{amsmath}
\usepackage{amsfonts}
\usepackage{amssymb}
\usepackage{epstopdf}
\usepackage{subfigure}
\usepackage{natbib}
\usepackage{color}%
\usepackage{hyperref}
\hypersetup{dvips,dvipdfm,linktoc=page,colorlinks=true,linkcolor=blue,citecolor=red,filecolor=magenta,urlcolor=magenta,bookmarks=true}
\providecommand{\U}[1]{\protect\rule{.1in}{.1in}}
\newcommand{\be}{\begin{equation}}
\newcommand{\ee}{\end{equation}}

\newcommand{\mincir}{\raise
-3.truept\hbox{\rlap{\hbox{$\sim$}}\raise4.truept\hbox{$<$}\ }}
\newcommand{\magcir}{\raise
-3.truept\hbox{\rlap{\hbox{$\sim$}}\raise4.truept\hbox{$>$}\ }}

\begin{document}
\title{ Interacting dark energy model in the brane scenario: A Dynamical System Analysis }
\author{Sujay Kr. Biswas}
\email{sujaymathju@gmail.com}
\affiliation{Department of Mathematics, Ramakrishna Mission Vivekananda Centenary College, Rahara, Kolkata-700 118, West Bengal, India.}
\affiliation{Department of Mathematics, Jadavpur University, Jadavpur, Kolkata - 700032, West Bengal, India}
\author{Subenoy Chakraborty}
\email{schakraborty@math.jdvu.ac.in}
\affiliation{Department of Mathematics, Jadavpur University, Jadavpur, Kolkata - 700032, West Bengal, India}
\keywords{brane world cosmology, dark energy, interaction, dynamical system,  phase space, stability}
\pacs{95.36.+x, 95.35.+d, 98.80.-k, 98.80.Cq.}
%%%%%%%%%%%%%%%%%%%%%%%%%%%%%%%%%%%%%%%%%%%%%%%%%%%%%%%%%%%%%%%%%%%%%%%%%%%%%%%%%%%%%%%%%%%%%%5
%%%%%%%%%%%%%%%%%%%%%%%%%%%%%%%%%%%%%%%%%%%%%%%%%%%%%%%%%%%%%%%%%%%%%%%%%%%%%%%%%%%%%%%%%%%%%%%%
\begin{abstract}
The present work is a second in the series of investigation of the background dynamics in brane cosmology when dark energy is coupled to dark matter by a suitable interaction. Here, dark matter is chosen in the form of perfect fluid with barotropic equation of state while a real scalar field with self interacting potential is chosen as dark energy. The scalar field potential is chosen as exponential or hyperbolic in nature and three different choices for the interaction between the dark species are considered. In the background of spatially flat, homogeneous and isotropic Friedmann-Robertson-Walker (FRW) brane model, the evolution equations are reduced to an autonomous system by suitable transformation of variables and a series of critical points are obtained for different interactions. By analyzing the critical points, we have found cosmologically viable model describing an early inflationary scenario to dark energy dominated era connecting through a matter dominated phase.

\end{abstract}
%%%%%%%%%%%%%%%%%%%%%%%%%%%%%%%%%%%%%%%%%%%%%%%%%%%%%%%%%%%%%%%%%%%%%%%%%%%%%%%%%%%%%%%%%%
%%%%%%%%%%%%%%%%%%%%%%%%%%%%%%%%%%%%%%%%%%%%%%%%%%%%%%%%%%%%%%%%%%%%%%%%%%%%%%%%%%%%%%%%%%
\maketitle
%%%%%%%%%%%%%%%%%%%%%%%%%%%%%%%%%%%%%%%%%%%%%%%%%%%%%%%%%%%%%%%%%%%%%%%%%%%%%%%%%%%%%%%%%%%%%%%%%%%%%%%%%%%%%%%%%%%%%%%%%%%%%%%%%%%%%%%%%%%%%%%%%%
\section{Introduction}

Brane world scenario \cite{Akama1,Randall1} is a modified theory of gravity
in which our four dimensional universe is embedded in a higher dimensional bulk space time.
A particular brane  scenario proposed by Randall and Sundrum \cite{Randall2} (known as RS2 brane)
consists of  a single positive tension brane as a sub-manifold in the five
dimensional anti- de Sitter ({\it i.e.}, $\Lambda=-\frac{6}{\ell^{2}}$, $ \ell $
is curvature radius of bulk) space ($ AdS_{5}$) with $ Z_{2}$  - symmetry
(the RS2 scenario \cite{Randall2}). In this modified gravity theory, the
Hubble parameter ($H$) is proportional to the energy density at early epochs
when $ \rho_{T} \gg \lambda $ (where $\lambda $ is the brane tension). However,
in the lower energy limit ({\it i.e.}, $ \rho_{T} \ll \lambda $) \cite{Randall2,Langlois1,Garriga1}
the standard four dimensional gravity is recovered.

Today, it is well known from various astrophysical observations \cite{Riess1,Davis1,MichaelWood1,Tegmark1,Jarosik1,Larson1,Komatsu1} that
presently the universe is going through an accelerating phase which can not be
explained by standard cosmology. So, Physicists are trying to resolve this
challenging issue either by modifying the right hand side (r.h.s) of the
Einstein field equations ({\it i.e.}, by introduction of some exotic matter
known as dark energy (DE)) or, by modifying the left hand side (l.h.s) of the
Einstein field equations ({\it i.e.}, introduction of modified gravity theory).
The unknown matter component, {\it i.e.}, DE is totally unknown to us except its
large negative pressure and is considered as an unresolved problem in modern
cosmology \cite{Sotiriou1,Capozziello1}. On the other hand, there are several
modified gravity theories, namely, $f(R)$-gravity, scalar-tensor gravity,
Einstein-Gauss-Bonnet gravity, Brane world gravity and so on \cite{Randall1,Randall2,Sotiriou1,Capozziello1,Nojiri1,Starobinsky1,Nojiri2,Nojiri3,Bengochea1}.
However, from the point of view of observational data, none of these models can fit
the observation in a better way compare to $\Lambda$CDM model, the simplest DE model
with cosmological constant $\Lambda$ representing the DE component, and the massive
non-relativistic particles in terms of Cold dark matter. Although, the $\Lambda$CDM
provides best fit to most of the recent observations, this model faces  some
severe problems such as, the cosmological constant problem \cite{T. Padmanabhan2003,S. Weinberg1989,V. Sahni2000}
and the coincidence problem \cite{I. Zlatev1999}. Dynamical dark energy models
(dark energy evolve with time), namely, quintessence, k-essence, tachyon, phantom and many others, are based on scalar field, have been studied in the literature to get a possible explanation for the cosmological constant problem.
However, uncoupled quintessence model
can not give rise to a solution of coincidence problem while a coupled quintessence
allows to give the possible solution of coincidence problem by providing scaling attractor \cite{C.G.Bohmer2008,C.Wetterich1995,L.Amendola1999}, since in the attractor regime,
both the dark components (DE and DM) scale in the similar way. However, it should be
noted that the models where DE interacts with the DM components of the universe,
were originally introduced as a solution to the$-$  coincidence problem.

Thus, the dynamics of  dark energy interacting with dark matter would provide better
results than the non-interacting one. Also, the argument behind choice of interaction
models is that they are favoured by observed data obtained from the Cosmic Microwave
Background (CMB) \cite{Oliveras1} and matter distribution at large scales \cite{Oliveras2}.
Further, Das et al \cite{Das1} and Amendola et al \cite{Amendola1} showed that an interaction
model of the universe mimics the observationally measured phantom equation of state as
compared to non-interacting models, which may predict a non phantom type of equation of state.
In fact, this interaction occurred non gravitationally is very weak or negligible in the sense
that DE is homogeneously distributed in the universe whereas DM clumps around the ordinary matter.
Since there is no fundamental theoretical approach for choosing such interaction, one assumes
this on complete phenomenological ground. Although, one may provide more physical and
natural results than others, so several interacting DE models are extensively studied in
the literature \cite{Yuri.L.Bolotin2014,Andre A.Costa2014,M.Khurshudyan2015,S.Kr.Biswas2015a,S.Kr.Biswas2015b,C.G.Bohmer2008,N.Tamanini2015,Xi-ming Chen2009,T.Harko2013,Nunes2016,Wang2016,Landim12016,Landim22016}.

Nevertheless, still people are trying to explain the observational fact by any one of
the above modified gravity theory or by imposing both type of modifications.
In the frame work of RSII brane scenario, several models have been proposed to
account for the present accelerating phase. In particular, a self interacting scalar field \cite{Maeda1,Pedro1,Majumdar1,Nunes1,Sami1} behaves as dark energy. The dynamics of
scalar field with constant or exponential \cite{Gonzalez1} as well as a wide variety
of self-interaction potentials \cite{Leyva1} have been studied in the context of FRW cosmology.
Recently, a detailed phase space analysis of RSII brane world scenario has been done in
Ref. \cite{Escobar1} and a complete cosmic scenario in this model has been shown
in \cite{Jibitesh1} with the help of dynamical system analysis. Also, scalar field
coupled to barotropic fluid has been studied in Refs. \cite{Copeland1,Aguirregabiria1,Lazkoz1,Fang1,Leon1,Leon2,Leon3}, and in this context, it is worthy to note that the relevant dynamical system studies in the brane
scenario can be found in \cite{Campos1,Campos2,Coley1,Coley2} and others (see for
instance \cite{Goheer1,Goheer2} where a phase space analysis with exponential potential
in the brane has been examined by Goheer and Dunsby ). Further, it should be noted in
this context that scalar field appears naturally in particle physics and in the present
context it behaves as a source of DE. For recent works on dynamical systems in the case of
modified gravity theories one may refer to references \cite{Odintsov2017,Oikonomou2018,Odintsov2018a,Kleidis2018,Odintsov2018b} and reference \cite{Bahamonde2018} for
a recent review on dynamical system.

The present work is continuation of our earlier work  in brane scenario \cite{S.Kr.Biswas2015a}.
Usually, brane world modified gravity theory is relevant only at early times whereas
interacting DE in GR can give a clear picture of evolution of late phase of universe.
Our main objective is to study interacting DE in context of brane scenario to achieve
a unified model of evolution of universe from early inflation to late time DE dominated
solution connected through a matter dominated era. Here, we have considered three sets
of new interaction term between perfect fluid having any barotropic equation of state and
a scalar field having self interacting potential as exponential or hyperbolic in form.
The evolution equations are converted to an autonomous system by suitable transformation
of the variables. A detail study of the critical points and their stability analysis
have been done in the context of cosmology. From the study, we have found some critical
points represent the scalar field dominated solutions which are accelerated attractors
in some parameter region but could not alleviate the coincidence problem. However,
the critical points representing the scaling attractors can solve the coincidence
problem successfully for some parameter restrictions when the potential of the self
interacting scalar field is taken as exponential. The paper is organized as follows :
Section \ref{formation_autonomous} comprises of the essential details of Randall - Sundrum model
and deals with basic equations in Brane Scenario and formation of Dynamical System.
In Section \ref{phase_space_auto}, a detailed phase space analysis related to the critical
points is presented. Also, we discuss the existence and stability / instability of critical
points for various interaction terms and potentials. Cosmological implications of critical
points for various system are shown in section \ref{cosmological_implications}. Finally,
the summary and concluding remarks are given in section \ref{summary}. Throughout the paper
we use natural units $\left(8\pi G=\frac{8 \pi}{m_{PL}^{2}}=\hbar=c=1  \right)$. \\

%%%%%%%%%%%%%%%%%%%%%%%%%%%%%%%%%%%%%%%%%%%%%%%%%%%%%%%%%%%%%%%%%%%%%%%%%%%%%%%%%%%%%%%%%%%%%%%%%%%%%%%%%%%%%%%%%%%%%%%%%%%

\section{ Formation of autonomous system in Brane scenario }
\label{formation_autonomous}

The present work deals with RSII brane scenario in the background of
homogeneous and isotropic spatially flat FLRW model of the universe.
The cosmic substratum is chosen as two interacting dark components$-$
one of the dark species, namely, the dark matter which is chosen as
perfect fluid with constant barotropic equation of state and the dark
energy component which is taken as real scalar field with arbitrary
self interaction potential. So the modified energy conservation equations
for the individual matter part take the form

\begin{equation}\label{continuity-matter}
    \dot{\rho_{m}} +3H\omega_{m}\rho_{m}=Q
\end{equation}
and
\begin{equation}\label{continuity-scalar}
    \dot{\rho_{\phi}} +3H(\rho_{\phi}+p_{\phi})=-Q.
\end{equation}
%%%%%%%%%%%%%%%%%%%%%%%%
The equation of state parameter for the matter field is
$W_{m}=p_m/\rho_m=\omega_{m}-1$ ( with $0 \leq \omega_{m} \leq 2 $),
where $\rho_{m}$ and $p_{m}$ represent the energy density and the
thermodynamic pressure for dark matter respectively. On the other hand,
the real scalar field $\phi$ represents the dark energy component with

\begin{equation}\label{Energydensity-scalar}
    \rho_{\phi}=\frac{1}{2}\dot{\phi}^{2} + V(\phi), ~~~~~ p_{\phi} = \frac{1}{2}\dot{\phi}^{2} - V(\phi)
\end{equation}
%%%%%%%%%%%%%%%%%%%%%%%%%%%%%%%%%%%%%%%
as the matter density and thermodynamic pressure. The scalar function $V(\phi)$
represents the self interaction potential for the scalar field. Now, using
(\ref{Energydensity-scalar}) in the conservation equation (\ref{continuity-scalar}),
we have the evolution equation for the scalar field as

\begin{equation}\label{EvolutionEqn-scalar}
    \ddot{\phi} + 3H\dot{\phi} + \frac{dV(\phi)}{d\phi}= -\frac{Q}{\dot{\phi}}.
\end{equation}

For the time being, the interaction term $Q$ is unspecified except that it is
assumed to have same sign throughout the cosmic evolution. In particular, $Q>0$
indicates a flow of energy from DE to DM and it is in the reverse direction for $Q<0$.
From physical point of view, $Q>0$, {\it i.e.} decaying of DE into DM, indicates the
validity of second law of thermodynamics. On the contrary, negative $Q$ ({\it i.e.},
decay of DM into DE) indicates the possibility that there is no DE field in the very
early universe and that DE `condenses' as a result of the slow decay
of DM \cite{Andre A.Costa2014}. Moreover, recently it has been shown \cite{C.G.Bohmer2008}
that the coupling parameter (in the interaction term) is weakly constrained to negative
values by Planck data set \cite{C.G.Bohmer2008}. Hence, it appears in the observed data
fittings that models with negative $Q$ show most significant departure from zero coupling.
Further, from the aspect of curvature perturbation, it is found \cite{Xi-ming Chen2009}
that there will be a stable curvature perturbation if
\begin{itemize}
\item (i)~$Q\propto \rho_{\phi}$ with $\omega_{\phi}=\frac{p_{\phi}}{\rho_{\phi}}\neq-1$, or
\item (ii)~$Q\propto \rho_{m}$, or $Q\propto \rho_{T}=\rho_{m}+\rho_{\phi}$ provided $\omega_{\phi}<-1.$
\end{itemize}
%%%%%%%%%%%%%%%%%%%
The modified Friedmann equations for the present brane scenario are given by \cite{Langlois2,Brax1,Langlois3,Maartens1}

\begin{equation}\label{Friedmann}
 H^{2} = { \frac{1}{3}}\rho_{T}(1+\frac{\rho_{T}}{2\lambda})+\frac{2U}{\lambda}
\end{equation}
and
\begin{equation}\label{Raychaudhuri}
  2 \dot{H} = - (1+\frac{\rho_{T}}{\lambda})(\dot{\phi}^{2}+\omega_{m}\rho_{m})-\frac{4U}{\lambda},
\end{equation}
%%%%%%%%%%%%%%%%%%%%%%%%%%%%%%%%%%%%%%%%%%%%%%%%%%%%%%%
where $ \lambda $ is the brane tension, $ \rho_{T}=\rho_{m} + \rho_{\phi} $
is the total matter energy density, $ U(t) =\frac{C}{a(t)^{4}} $ ($a$ is the
scale factor for the FRW model) is the dark radiation due to non- zero bulk Weyl tensor,
and the constant parameter $C$ is  related to the black hole mass in the bulk.
In particular, for AdS- Schwarzschild bulk  $ C $ to be non-zero \cite{Brax1},
but C identically vanishes \cite{Maartens1,Bowcock1}  for AdS bulk.
For simplicity of calculation, we shall restrict ourselves to AdS bulk model
so that $ C $ is chosen to be zero. Note that for small brane tension
($\lambda$), {\it i.e.}, $ \rho_{T} \gg \lambda $, the brane effect will be
significant ({\it i.e.}, $ H \propto \rho_{T} $) at early phases of evolution
of universe while at the late phase of evolution ({\it i.e.}, $\rho_{T} \ll \lambda$)
the brane effects will be not so significant ($ H \propto \sqrt{\rho_{T}} $).
In any case, the modified Friedmann equations can describe the evolution of
the universe at all times.

The evolution equations, namely, (\ref{continuity-matter}),  (\ref{EvolutionEqn-scalar}),
(\ref{Friedmann}), and (\ref{Raychaudhuri}) in the brane scenario are
highly non-linear and coupled second order differential equations with
$ a$, $\rho_{m}$, and $\phi $ as the dependent variables and the cosmic
time `t' is the independent variable. (Note that equation (\ref{Friedmann})
may be considered as the constraint equation related to the dependent variables).
As it is not possible to have an analytic solution describing the evolution of the
universe so for a qualitative idea about the cosmic evolution we shall write the
evolution equations in an autonomous dynamical system. As a first step, we introduce
the following variables\cite{Escobar1}:

\begin{equation}\label{Variables}
    x=\frac{\dot{\phi}}{\sqrt{6}H},~~  y= \frac{V}{3H^{2}},~~  z= \frac{\rho_{T}^{2}}{6\lambda H^{2}},
\end{equation}
%%%%%%%%%%%%%%%%%%%%%%%%%%%%%%%%%%%%%%%%%%%%%%%%%%%%%%%
which are normalized over the Hubble scale. As a result, the evolution equations
reduce (after some algebra) to the following autonomous system of ordinary differential equations:
%%%%%%%%%%%%%%%%%%%%%%%%%%%%%%%%%%%%%%%%%%%%%%%%%%%%%%%%%%%%%%%%%%%%%%%%%
\begin{eqnarray}
\begin{split}
   \frac{dx}{dN}& = \sqrt{\frac{3}{2}}ys - 3x + \frac{3}{2}x^{3} \frac{(1+z)}{(1-z)} (2-\omega_{m})
   + \frac{3}{2}x\omega_{m}(1-y-z) \frac{(1+z)}{(1-z)} -\frac{Q}{6xH^{3}},& \\
   \frac{dy}{dN}& = -\sqrt{6} xys + 3y\frac{(1+z)}{(1-z)}\left[x^{2}(2-\omega_{m})+\omega_{m}(1-y-z)\right],& \\
   \frac{dz}{dN} & = -3z\left[x^{2}(2-\omega_{m}) +\omega_{m} (1-y-z)\right],& \\
    \frac{ds}{dN}& = -\sqrt{6} x f(s).&~~\label{autonomousGen}
\end{split}
\end{eqnarray}
%%%%%%%%%%%%%%%%%%%%%%%%%%%%%%%%%%%%%%%%%%%%%%%%%%%%%%%%%%%%%%%
Here, $ N=\ln a $ is chosen as the independent variable.
Dynamical variable $``s"$ is defined as

\begin{equation}\label{s-variable}
    s=-\frac{V'}{V},
\end{equation}
%%%%%%%%%%%%%%%%%%%%%%%%%%%%%%%%%%
where $V'\equiv \frac{dV}{d\phi}$ and we have

\begin{equation}
    f(s)=\frac{V''}{V}-s^{2}.
\end{equation}\\
%%%%%%%%%%%%%%%%%%%%%%%%%%%%%%%%%%%%%%%%%%%

Thus the system of equations in (\ref{autonomousGen}) form an autonomous system
$\overrightarrow{\alpha}=f(\overrightarrow{\alpha})$ with $\overrightarrow{\alpha}=\left(
                                                                                     \begin{array}{c}
                                                                                       x \\
                                                                                       y \\
                                                                                       z \\
                                                                                       s \\
                                                                                     \end{array}
                                                                                   \right),$
provided the potential function $V(\phi)$   is chosen such that $\frac{V^{''}}{V}$ can be
expressed as a function of $\frac{V^{'}}{V}$, {\it i.e.}, $``s"$. In the present work, the potential function
is chosen as
$$(i)~V=V_{0}exp(-\mu \phi),~~ \mbox{or} ~~(ii)~V=V_{0}\cosh (\mu \phi)$$
%%%%%%%%%%%%%%%%%%%%%%%%%%%%%%%%%%%%%%%%%%%%%%%%%%%%%%%%%%%%%%
with $V_{0}$ and $\mu $ as constant. Note that, for the exponential potential
the variable $``s"$ turns out to be a constant ($\mu$) so the autonomous system
reduces to three dimensional phase space, while for the hyperbolic potential
$f(s)=\mu^{2}-s^{2}$.
In the following section, we shall explicitly analyze the autonomous
system (Eq.(8)) for three different choices for the interaction term, namely,
$(i)~Q1=\gamma \dot{\phi}\rho_{m}$\cite{C.Wetterich1995,L.Amendola1999,Kaeonikhom1},~~
$(ii)~Q2=\delta \frac{\rho_{m}}{H}\dot{\phi}^{2}$\cite{Xi-ming Chen2009},~~and~~
$(iii)~Q3=\sigma \frac{\rho_{m}^{2}}{H}$\cite{Xi-ming Chen2009}
%%%%%%%%%%%%%%%%%%%%%%%%%%%%%%%%%%%%%
with $\gamma$, $\delta$, and $\sigma$ as the coupling parameters.
It should be noted that there are no
guiding principles in choosing the forms of couplings and the potentials that are
studied in the paper, rather they are chosen more for the
convenience to perform dynamical analysis. Also the potentials
and couplings are expected to give interesting physical results.
\\

%%%%%%%%%%%%%%%%%%%%%%%%%%%%%%%%%%%%%%%%%%%%%%%%%
Using the normalized variables (in equation (\ref{Variables})) into the first
modified Friedmann equation ({\it i.e.}, equation (\ref{Friedmann})), one obtains the
density parameter for dark matter as

 \begin{equation}\label{DensityDM}
    \Omega_{m}=1-x^{2} -y-z
\end{equation}
%%%%%%%%%%%%%%%%%%%
As $ 0\leq \Omega_{m} \leq 1 $, so the normalized variables are not independent,
rather they are constrained as

\begin{equation}
    0\leq x^{2} +y + z \leq 1.
\end{equation}
%%%%%%%%%%%%%%%%%%%%%%%%%%%%%%%%%%%%%%%%%%
Also, the ratio of the total energy density to the brane tension is given by

\begin{equation}
    \frac{\rho_{T}}{\lambda} = \frac{2z}{(1-z)}.
\end{equation}
%%%%%%%%%%%%%%%%%%%%%%%%%%%%%%%%%%%%%%%%%%%%%%%
From the above relation, it is clear that the early super dense region
$(\rho_{T}\gg \lambda)$ is confined to the neighbourhood of  $z=1$ which
represents the initial big-bang singularity. Similarly, the late phase of
the evolution is represented in the neighborhood of $z=0$. However, $z=1$
is not allowed by the above autonomous system, {\it i.e.}, our model is not
appropriate to describe the evolution  dynamics near the initial big-bang
singularity (possibly due to quantum effects). But, from the mathematical
point of view one may obtain the neighborhood of this initial singularity
in the limiting sense ({\it i.e.}, asymptotically). Thus, the phase space of
the above autonomous system can be described as

\begin{equation}\label{phase-boundary-EXP}
    \Omega_{\rho s} = \left[\{ x,y,z \} \times \{ s \}: 0\leq x^{2} + y +z \leq 1,~~ |x|\leq 1,~~ 0\leq y \leq 1,~~ 0 \leq z \leq 1,~~ s\in \mathbb{R} \right].
\end{equation}
%%%%%%%%%%%%%%%%%%%%%%%%%%%%%%%%%%%%%%%%%%%%%%%%
 Also, the other relevant cosmological parameters in terms of the new variables take the form :

\begin{equation}
    \omega_{\phi} = \frac{p_{\phi}}{\rho_{\phi}} =\frac{x^{2}-y}{x^{2}+y},~~ \Omega_{\phi} = \frac{\rho_{\phi}}{3H^{2}} = x^{2} + y,
\end{equation}
%%%%%%%%%%%%%%%%%%%%%%%%%%%%%%%%%%%%%%%%%%%%%%%%%%%%%%%%%
the deceleration parameter takes the form
\begin{equation}\label{acceleration}
    q= -1-\frac{\dot{H}}{H^{2}}= -1 + \frac{3}{2} \left(\frac{1+z}{1-z} \right) \left[x^{2}(2-\omega_{m})+\omega_{m}(1-y-z) \right],
\end{equation}
%%%%%%%%%%%%%%%%%%%%%%%%%%%%%%%%%%%%%%%%%%%%%%%%%%%%%%%%
 and the effective equation of state reads as
\begin{equation}\label{effective-eos}
\omega_{eff}=-1+\omega_{m}+\frac{x^{2}(2-\omega_{m})-\omega_{m}y}{1-z}
\end{equation}

%%%%%%%%%%%%%%%%%%%%%%%%%%%%%%%%%%%%%%%%%%%%%%%%%%%%%%%%%%%%%%%%%%%%%%%%%%%%%%

\section{Phase space analysis of autonomous system (\ref{autonomousGen}) for various choices of interaction and potential:}
\label{phase_space_auto}

We shall now discuss the phase space analysis of interacting DE with various
form of interaction term. The potential of the associated scalar field is chosen
as exponential or hyperbolic in nature. Critical points will be analyzed for the
different autonomous systems provided by the interactions. Then, linear stability
analysis will be done for different cases step by step.

\subsection{Interaction Model 1 Q1: $Q=\gamma \dot{\phi}\rho_{m}$ }\label{Interaction1}

First, we consider the interaction term as :

\begin{equation}\label{Interaction1}
Q=\gamma \dot{\phi}\rho_{m}.
\end{equation}
For this interaction, we first consider the exponential potential

\subsubsection{Exponential potential: $V=V_{0} exp(-s\phi)$ }

 Using the above interaction term of equation (\ref{Interaction1}), the autonomous
 system (\ref{autonomousGen}) with this exponential potential takes the form:
%%%%%%%%%%%%%%%%%%%%%%%%%%%%%%%%%%%%%%%%%%%%%%%%%%%%%%%%%%
\begin{eqnarray}
\begin{split}
   \frac{dx}{dN}& = \sqrt{\frac{3}{2}}ys - 3x + \frac{3}{2}x^{3} \frac{(1+z)}{(1-z)} (2-\omega_{m})+ \frac{3}{2}x\omega_{m}(1-y-z) \frac{(1+z)}{(1-z)} -\sqrt{\frac{3}{2}}\gamma(1-x^{2}-y-z),& \\
   \frac{dy}{dN}& = -\sqrt{6} xys + 3y\frac{(1+z)}{(1-z)}\left[x^{2}(2-\omega_{m}) + \omega_{m} (1-y-z)\right],&\\
   \frac{dz}{dN} &= -3z\left[x^{2}(2-\omega_{m}) +\omega_{m} (1-y-z)\right].&~\label{auto-Exp-Int-1}
\end{split}
\end{eqnarray}
%%%%%%%%%%%%%%%%%%%%%%%%%%%%%%%%%%%%%%%%%%%%%%%%%%%%%%%%%%%
Note that, for exponential potential $s(=\mu)$ is a constant.
The critical points for the system (\ref{auto-Exp-Int-1}) are the following
\begin{itemize}
\item  I. Critical Points : $ A1,~ A2 = ( \pm1, 0,0  ) $
\item  II. Critical Point : $ A3 = \left( \frac{\sqrt{6}}{3}\frac{\gamma}{(-2+\omega_{m})},0,0 \right ) $
\item  III. Critical Point : $ A4 = \left( \frac{s}{\sqrt{6}} , 1-\frac{s^{2}}{6},0 \right) $
\item  IV. Critical Point : $ A5= \left (\frac{\sqrt{6}}{2}\frac{\omega_{m}}{(\gamma+s)}, \frac{\gamma}{\gamma+s}+\frac{3}{2}\frac{\omega_{m}(2-\omega_{m})}{(\gamma+s)^{2}},0  \right) $

\item  V. Critical Point : $ A6= \left ( \frac{\sqrt{6}}{2}\frac{\omega_{m}}{\gamma} ,0, \frac{1}{2}\frac{2\gamma^{2}-3\omega_{m}^{2}+6\omega_{m}}{\gamma^{2}}  \right).$
\end{itemize}
%%%%%%%%%%%%%%%%%%%%%%%%%%%%%%%
Critical points and their corresponding physical parameters are shown in the table \ref{modelQ1EXP} and the eigenvalues of linearized Jacobian matrix at the critical points are shown in the tabular form (see in the table \ref{modelQ1EXP-Eigen}).
%%%%%%%%%%%%%%%%%%%%%%%%%%%%%%%%%%%%%%%%%%%%%%%%%%%%%%%%%%%
%TCIMACRO{\TeXButton{B}{\begin{table}[tbp] \centering}}%
%BeginExpansion
\begin{table}[tbp] \centering
%EndExpansion
\caption{The Critical Points and the corresponding physical parameters for the interaction model $ Q1=\gamma \dot{\phi}\rho_{m}$ for exponential potential are presented.}%
\begin{tabular}
[c]{ccccccc}\hline\hline
\textbf{Critical Points}&$(\mathbf{x,y,z})$& $\mathbf{\omega_{\phi}}$ &
 $\mathbf{\Omega_{m}}$ & $\mathbf\Omega_{\phi}$ &
\textbf{q}\\\hline
$A1,A2  $ & $( \pm1, 0,0  )$&  $ 1 $ &
$0$ & $1$ & $2$\\
$A3  $ & $\left( \frac{\sqrt{6}}{3}\frac{\gamma}{(-2+\omega_{m})},0,0 \right )$ & $ 1  $ &
 $1-\frac{2}{3}\frac{\gamma^{2}}{(-2+\omega_{m})^{2}}$ & $ \frac{2}{3}\frac{\gamma^{2}}{(-2+\omega_{m})^{2}} $ & $ \frac{8\omega_{m}-4+2\gamma^{2}-3\omega_{m}^{2}}{2(2-\omega_{m})} $ \\
$A4  $ & $ \left( \frac{s}{\sqrt{6}} , 1-\frac{s^{2}}{6},0 \right)$& $ \frac{1}{3}s^{2}-1  $ &
 $ 0 $ & $ 1 $ & $ -1+\frac{1}{2}s^{2} $ \\
 $A5  $ & $ \left (\frac{\sqrt{6}}{2}\frac{\omega_{m}}{(\gamma+s)}, \frac{\gamma}{\gamma+s}+\frac{3}{2}\frac{\omega_{m}(2-\omega_{m})}{(\gamma+s)^{2}},0  \right)$& $ \frac{3\omega_{m}^{2}-\gamma^{2}-\gamma s-3 \omega_{m}}{\gamma^{2}+\gamma s+3\omega_{m}}  $ &
 $ \frac{\gamma s+s^{2}-3\omega_{m}}{(\gamma+s)^{2}} $ & $ \frac{\gamma^{2}+\gamma s+3\omega_{m}}{(\gamma+s)^{2}} $ & $ -1+\frac{3}{2}\frac{\omega_{m}s}{(\gamma+s)}$ \\

 $A6  $ & $\left ( \frac{\sqrt{6}}{2}\frac{\omega_{m}}{\gamma} ,0, \frac{1}{2}\frac{2\gamma^{2}-3\omega_{m}^{2}+6\omega_{m}}{\gamma^{2}}  \right) $& $ 1 $ &
 $ -\frac{3\omega_{m}}{\gamma^{2}} $ & $ \frac{3}{2}\frac{\omega_{m}^{2}}{\gamma^{2}} $ & $ -1$
 \\\hline\hline
\end{tabular}
\label{modelQ1EXP}
%TCIMACRO{\TeXButton{E}{\end{table}}}%
%BeginExpansion
\end{table}%
%EndExpansion
%
%%%%%%%%%%%%%%%%%%%%%%%%%%%%%%%%%%%%%%%
%TCIMACRO{\TeXButton{B}{\begin{table}[tbp] \centering}}%
%BeginExpansion
\begin{table}[tbp] \centering
%EndExpansion
\caption{The eigenvalues of the linearized system  for the interaction model $ Q1=\gamma \dot{\phi}\rho_{m}$ for exponential potential, where $\Delta_{A5}=\sqrt{36 s^{2}-180s^{2}\omega_{m}+72\gamma s+81\omega_{m}^2 s^{2}-36\gamma s\omega_{m}+36\gamma^{2}-48\gamma^{3} s-216\omega_{m}^{3}+144\omega_{m} \gamma^{2}-96 s^{2}\gamma^{2}-48 s^3 \gamma+72\omega_{m}^{2} \gamma s+432\omega_{m}^{2}}$}%
\begin{tabular}
[c]{ccccccc}\hline\hline
\textbf{Critical Points} & $\mathbf{\lambda_{1}}$ &
 $\mathbf{\lambda_{2}}$ & $\mathbf{\lambda_{3}}$ \\\hline
$A1  $ & $ -6 $ &  $ 6-3\omega_{m}+\gamma \sqrt{6}$ & $ 6-s \sqrt{6}$ \\
$A2  $ & $ -6 $ &  $ 6+s \sqrt{6} $ & $ 6-3\omega_{m}-\gamma \sqrt{6}$ \\
$A3  $ & $  \frac{3\omega_{m}^{2}-6\omega_{m}-2\gamma^{2}}{2-\omega_{m}}  $ &
 $ \frac{2\gamma^{2}-3\omega_{m}^{2}+12\omega_{m}-12}{2(2-\omega_{m})} $ & $ \frac{2\gamma^{2}-3\omega_{m}^{2}+6\omega_{m}+2\gamma s}{(2-\omega_{m})} $  \\
$A4  $ & $ -s^{2} $ & $ \frac{1}{2}s^{2}-3  $ &
 $ \gamma s+s^{2}-3\omega_{m} $  \\
 $A5  $ & $ -\frac{3\omega_{m} s}{\gamma+s} $ & $ \frac{1}{(\gamma+s)}(-\frac{3s}{2}+\frac{3s \omega_{m}}{4}-\frac{3\gamma}{2}+\frac{\Delta_{A5}}{4})  $ &
 $ \frac{1}{(\gamma+s)}(-\frac{3s}{2}+\frac{3s \omega_{m}}{4}-\frac{3\gamma}{2}-\frac{\Delta_{A5}}{4}) $ \\

 $A6  $ & $ -\frac{3\omega_{m} s}{\gamma} $ & $ -\frac{3}{2}\frac{\gamma-\sqrt{\gamma^{2}+6\omega_{m}^{3}-12\omega_{m}^{2}-4\omega_{m} \gamma^{2}}}{\gamma} $ &
 $ -\frac{3}{2}\frac{\gamma+\sqrt{\gamma^{2}+6\omega_{m}^{3}-12\omega_{m}^{2}-4\omega_{m} \gamma^{2}}}{\gamma} $
 \\\hline\hline
\end{tabular}
\label{modelQ1EXP-Eigen}
%TCIMACRO{\TeXButton{E}{\end{table}}}%
%BeginExpansion
\end{table}%
%EndExpansion
%%%%%%%%%%%%%%%%%%%%%%%%%%%%%%%%%%%%%%%%%%%%%%%%%%%%%%%%

$\bullet $  The critical points A1 and A2 always exist in the phase space.
These solutions dominated by the kinetic part of the scalar field
correspond to standard cosmological behavior ({\it i.e.}, $z=0$). The DE behaves
as stiff fluid in this case. There  exists always a decelerating phase near the points.
They are hyperbolic in nature (see table \ref{modelQ1EXP-Eigen}). The point A1 admits a stable solution for the
restrictions  $s>\sqrt{6}$ and $\gamma<-\sqrt{\frac{3}{2}}(2-\omega_{m})$, while
the point A2 shows the stable solution for $s<-\sqrt{6}$ and $\gamma>\sqrt{\frac{3}{2}}(2-\omega_{m})$.
Thus, these two stable hyperbolic critical points indicate late time decelerating phase of
the universe. Only future observations will predict whether they are realistic or not.

$\bullet $ The solution represented by the point A3 is a combination of both DE and DM.
It will exist for $0\leq \omega_{m} <2$ and $\frac{3 \omega_{m} -6}{\sqrt{6}}\leq \gamma \leq \frac{6-3 \omega_{m} }{\sqrt{6}}$. Here, DE behaves as stiff fluid ($\omega_{\phi}=1$). A DM dominated universe near the point can be obtained for uncoupled case (i.e., when $\gamma=0$). There exists an accelerating universe
near the point when $-\sqrt{2}<\gamma <\sqrt{2}$ and $0\leq \omega_{m} <\frac{4}{3}-\frac{1}{3} \sqrt{2} \sqrt{3 \gamma ^2+2}$. The point is stable for either (i) $-\sqrt{6}<\gamma<0$, $0\leq\omega_{m}<\frac{2\sqrt{6}+2\gamma}{\sqrt{6}}$ and
$s>\frac{-2\gamma^{2}-6\omega_{m}+3\omega_{m}^{2}}{2\gamma}$ or (ii) $0<\gamma<\sqrt{6}$, $0\leq\omega_{m}<\frac{2\sqrt{6}-2\gamma}{\sqrt{6}}$ and
$s<\frac{-2\gamma^{2}-6\omega_{m}+3\omega_{m}^{2}}{2\gamma}$. Thus, the critical point has
a significant role in the late time dynamics of the universe.

$\bullet $  Completely scalar field dominated solution A4 will exist for
$s^{2}\leq6$. Here, DM is absent and any perfect fluid represents the DE with
equation of state parameter $\omega_{\phi}=\frac{s^{2}}{3}-1$. There exists an
accelerating phase of the universe near this point for $ s^{2}<2 $. From the linear
stability analysis, we observe (see table \ref{modelQ1EXP-Eigen}) that the conditions
for stability of scalar field dominated hyperbolic critical point are:

$\gamma \in \mathbb{R}~~\mbox{and ~either}\\
 (i)\left\{ -\sqrt{6}<s<0,~~~~ 0\leq \omega_{m} \leq 2,~~\mbox{and}~~ \gamma >\frac{3 \omega_{m} -s^2}{s}\right\},~~\mbox{or}~~\\
(ii) \left\{s=0,~~\mbox{and}~~ 0<\omega_{m} \leq 2 \right\},~~\mbox{or}~~\\
(iii) \left\{0<s<\sqrt{6},~~~~ 0\leq \omega_{m} \leq 2,~~\mbox{and}~~ \gamma <\frac{3 \omega_{m} -s^2}{s} \right \}$.\\

\begin{figure}
\centering
\includegraphics[width=7cm,height=5cm]{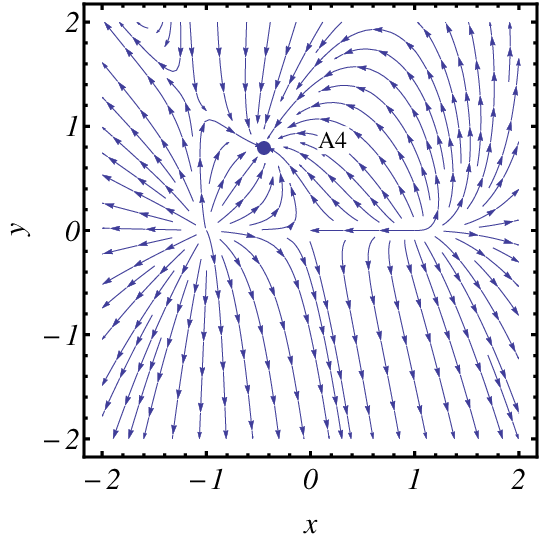}
\caption{Vector field of autonomous system (\ref{auto-Exp-Int-1}) with the interaction (\ref{Interaction1}) shows a stable attractor A4 for the parameters values $\gamma =0.01$, $s=-1.1$, $\omega_{m} =1.01.$}
\label{phasespace-coupling1-fig1}
\end{figure}

\begin{figure}
\centering
\includegraphics[width=7cm,height=5cm]{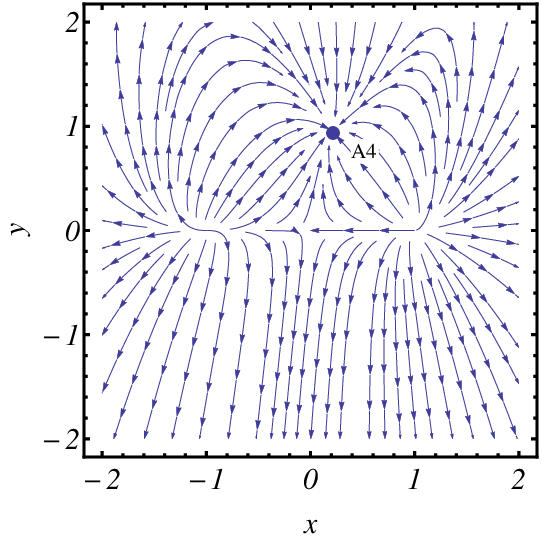}
\caption{Figure shows projection of vector field of autonomous system (\ref{auto-Exp-Int-1}) in the $x-y$ plane. The critical points A4 exhibits a stable solution for the parameter values $\gamma =0.01$, $s=0.5$, $\omega_{m} =1.01$.}
\label{phasespace-coupling1-fig2}
\end{figure}

In figure (\ref{phasespace-coupling1-fig1}), for the independent parameters
$\gamma =0.01$, $s=-1.1$, and $\omega_{m} =1.01$, the point A4 represents the
stable solution and another figure (\ref{phasespace-coupling1-fig2}),
for $\gamma =0.01$, $s=0.5$, $\omega_{m} =1.01$, shows that the point A4
is stable solution. Note that the point recovers $\Lambda$CDM for constant
potential ({\it i.e.}, for $s=0$, see table \ref{modelQ1EXP}).
The point does not solve the coincidence problem (since $\Omega_{\phi}=1$).
This stable critical point represents the late time accelerating phase of the
universe and it is fully dominated by DE.
%%%%%%%%%%%%%%%%%%%%%%%%%%%%%%%%%%%%%%%%%%%%%%%%%%%%%%%

%%%%%%%%%%%%%%%%%%%%%%%%%%%%%%%%%%%%%%%%%%%%%%%%%%%%%%%%
$\bullet $ The solution A5 is the combination of both the DE and DM components.
It exists for

$(1)~~\omega_{m} =0: ~~\mbox{and}~ \mbox{either}~ \mbox{of} $\\
$(a)s<0~~ \mbox{and} ~~\gamma <-s, ~~\mbox{or} $\\
$(b)s=0 ,~~ \mbox{and}~~ ( \gamma <0,~~ \mbox{or}~~ \gamma >0 ),~~ \mbox{or} $\\
$(c)s>0,~~ \mbox{and} ~~\gamma >-s $\\

$(2)~~0<\omega_{m} <2:~~~~\mbox{and}~ \mbox{either}~ \mbox{of} $\\
$(a)~s\leq -\sqrt{6},~~\mbox{and}~~ \gamma \leq \frac{1}{2} \left(-2 s-\sqrt{6} \omega_{m} \right)$\\
$(b)~-\sqrt{6}<s<0,~~\mbox{and}~~ \gamma \leq \frac{3 \omega_{m} -s^2}{s}$\\
$(c)~0<s\leq \sqrt{6}~~\mbox{and}~~ \gamma \geq \frac{3 \omega_{m} -s^2}{s}$\\
$(d)~s>\sqrt{6},~~\mbox{and}~~ \gamma \geq \frac{1}{2} \left(\sqrt{6} \omega_{m} -2 s\right)$\\

$(3)~~\omega_{m} =2: ~~\mbox{and}~ \mbox{either}~ \mbox{of} $\\
$(a)~s<-\sqrt{6},~~\mbox{and}~~ \gamma \leq \frac{1}{2} \left(-2 s-2 \sqrt{6}\right)$ \\
$(b)~-\sqrt{6}\leq s<0,~~\mbox{and}~~ \gamma \leq \frac{6-s^2}{s}$\\
$(c)~0<s\leq \sqrt{6},~~\mbox{and}~~ \gamma \geq \frac{6-s^2}{s}$\\
$(d)~s>\sqrt{6},~~\mbox{and} ~~\gamma \geq \frac{1}{2} \left(2 \sqrt{6}-2 s\right).$\\
%%%%%%%%%%%%%%%%%%%%%%%%%%%%%%%%%%%%%%%%%%%%%%%%%%%%%%
Expansion of the universe near the critical point is accelerating for
$\omega_{m}<\frac{2}{3}(\frac{\gamma}{s}+1)$ (see table \ref{modelQ1EXP}).
The conditions for stability of the hyperbolic (see table \ref{modelQ1EXP-Eigen}), scaling solution (matter-scalar field) for the critical  point A5 are:\\

$(2)~~0<\omega_{m} <2: $\\
$(a)~s\leq -\sqrt{6},~~\mbox{and}~~ \gamma \leq \frac{1}{2} \left(-2 s-\sqrt{6} \omega_{m} \right)$\\
$(b)~-\sqrt{6}<s<0,~~\mbox{and}~~ \gamma \leq \frac{3 \omega_{m} -s^2}{s}$\\
$(c)~0<s\leq \sqrt{6}~~\mbox{and}~~ \gamma \geq \frac{3 \omega_{m} -s^2}{s}$\\
$(d)~s>\sqrt{6},~~\mbox{and}~~ \gamma \geq \frac{1}{2} \left(\sqrt{6} \omega_{m} -2 s\right)$\\

\begin{figure}
\centering
\includegraphics[width=7cm,height=5cm]{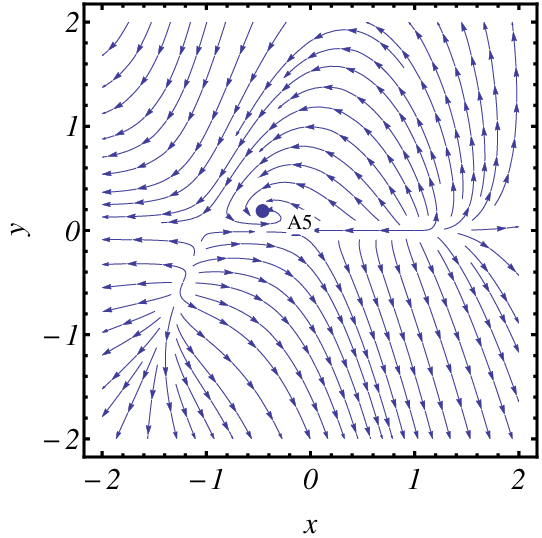}
\caption{Phase portrait of autonomous system (\ref{auto-Exp-Int-1}) in  the $x-y$ plane shows that the point A5 is stable solution for $\gamma =0.01$, $s=-3.0$, $ \omega_{m} =1.01.$}
\label{phasespace-coupling1-fig3}
\end{figure}

\begin{figure}
\centering
\includegraphics[width=7cm,height=5cm]{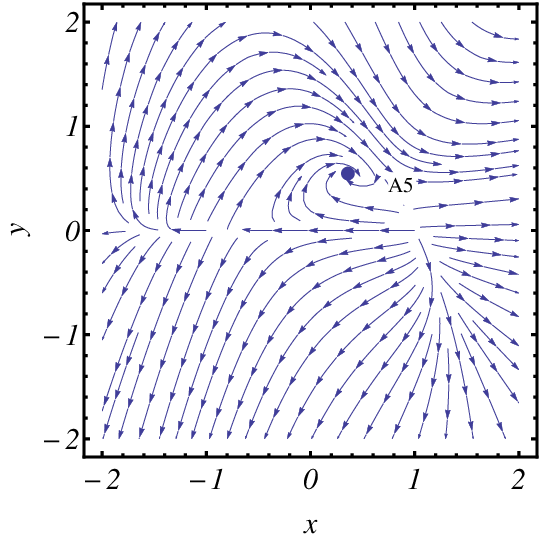}
\caption{The point A5 is stable solution for the parameters values $\gamma =1.1$, $s=2$, $\omega_{m} =1. $}
\label{phasespace-coupling1-fig4}
\end{figure}

\begin{figure}
\centering
\includegraphics[width=7cm,height=5cm]{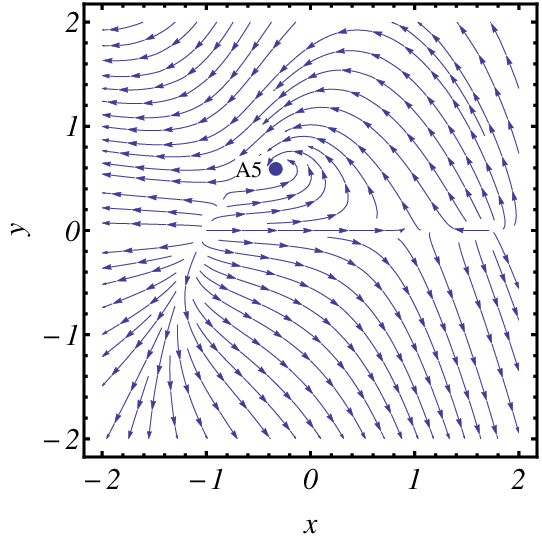}
\caption{The figure shows the phase portrait of autonomous system (\ref{auto-Exp-Int-1}) with interaction  (\ref{Interaction1}) for the parameters values $ \gamma =-2.1$, $s=-1.9$, $\omega_{m} =1$, where A5 is stable solution in x-y plane.}
\label{phasespace-coupling1-fig5}
\end{figure}

$(3)~~\omega_{m} =2: $\\
$(a)~s<-\sqrt{6},~~\mbox{and}~~ \gamma \leq \frac{1}{2} \left(-2 s-2 \sqrt{6}\right)$ \\
$(b)~-\sqrt{6}\leq s<0,~~\mbox{and}~~ \gamma \leq \frac{6-s^2}{s}$\\
$(c)~0<s\leq \sqrt{6},~~\mbox{and}~~ \gamma \geq \frac{6-s^2}{s}$\\
$(d)~s>\sqrt{6},~~\mbox{and}~~ \gamma \geq \frac{1}{2} \left(2 \sqrt{6}-2 s\right).$\\
%%%%%%%%%%%%%%%%%%%%%%%%%%%%%%%%%%%%%%%%%%%%%%%%%%%%%%%%%%%%%%%%%%%%%%%%%%%%%%%%%%%%%%%%%
The matter-scalar field scaling solution A5 is a accelerated stable
attractor which is shown in the figures. For example, in figure
(\ref{phasespace-coupling1-fig3}), for the parameters values
$\gamma =0.01$, $s=-3.0$, $\omega_{m} =1.01$, the point A5 represents
the stable attractor. For $\gamma =1.1$, $s=2$, $\omega_{m} =1$,
the figure (\ref{phasespace-coupling1-fig4}) shows that the point
A5 scaling attractor, and in the figure (\ref{phasespace-coupling1-fig5})
with the parameter values $\gamma =-2.1$, $s=-1.9$, $\omega_{m} =1$,
the point A5 is stable attractor (with $\Omega_\phi=0.71$, $\Omega_m=0.29$, and $\omega_{eff}=-0.53$).
Thus, this critical point represents
a scaling solution and corresponds to late time acceleration.
 It should be mentioned that for $\omega_{m}=0$
(i.e., matter is like cosmological constant), the scaling solution $A5$
becomes nonhyperbolic type critical point and then, it has a 2D stable sub manifold either
in the region $\gamma>0$, $s>0$ or $\gamma<0$, $s<0$. For this case,
the ratio of $\Omega_{m}$ and $\Omega_{\phi}$ takes the form $\Omega_{m}/\Omega_{\phi}=s/\gamma$
which matches the observation. Therefore, the point $A5$ would behave as late time de Sitter like
solution ($\omega_{eff}=-1$) for the interacting two fluid model in brane scenario.
%%%%%%%%%%%%%%%%%%%%%%%%%%%%%%%%%%%%%%%%%%%%%%%%%%%%%%%%%%

%%%%%%%%%%%%%%%%%%%%%%%%%%%%%%%%%%%%%%%%%%%%%%%%%%%%%%%%%%%%%%%%%%%%%%%%%

$\bullet $  The point A6 is the solution with 5D corrections in RSII brane scenario.
This point exists either for (i) $\omega_{m}=0$, ($\gamma<0$, or $\gamma>0$), or for
(ii) $\omega_{m}<0$, ($\gamma\leq \sqrt{\frac{3}{2}}\omega_{m}$ or $\gamma\geq -\sqrt{\frac{3}{2}}\omega_{m}$).
It represents the early Big Bang singularity (as $z\rightarrow 1$, $\rho_{T}\rightarrow \infty$) due to the existence criteria ($\omega_{m}=0$). Here, DE behaves as stiff fluid. There exists always an accelerating phase ($q=-1$) near the point. For the condition, $\omega_{m}=0$ the two eigenvalues become zero, and as a result, the point is a non-hyperbolic one (see table \ref{modelQ1EXP-Eigen}) with 1D stable sub-manifold. Here, the universe is dominated by the energy density (total) fully contributed by the brane effects only in its early evolution where $\Omega_{\phi}=0$, $\Omega_{m}=0$, $z=1$.
 The point also exists for $\omega_{m}<0$ (here, matter is like phantom fluid) and $\gamma<0$
describing an unstable (saddle) solution in the phase space
and in this case, an early inflationary scenario can be observed.

%%%%%%%%%%%%%%%%%%%%%%%%%%%%%%%%%%%%%%%%%%%%%%%%%%%%%%%%%%%%%%%%%%%%%%%%%%%%%%%%%%%%

\subsubsection{Hyperbolic potential: $V=V_{0} \cosh(\mu\phi)$ }

If we consider a wider class of self-interaction potentials beyond the constant
and exponentials the system (\ref{autonomousGen}) is not a closed system of
ordinary differential equations as in general $\frac{V'}{V}$ defined
in (\ref{s-variable}) is a function of scalar field itself. In order to go
further to analyze the phase space dynamics of the system one has to take
account of an extra variable $s$ in the autonomous system (\ref{autonomousGen}).
Thus, we introduce a new dynamical variable for studying the dynamics under
hyperbolic potential $V=V_{0} \cosh(\mu\phi)$\cite{Sahni1,Sahni2,Salcedo1}.
The extra equation will be in the form:

\begin{equation}\label{s-equation}
\frac{ds}{dN} = -\sqrt{6} xf(s)
\end{equation}
%%%%%%%%%%%%%%%%%%%%%%%%%%%%%%%%%%%%%%%%%%%%%%%%%%
where $f(s)=s^{2}\Gamma-s^2$, with the quantity $\Gamma=\Gamma(\phi) =\frac{V''V}{V'^{2}}$
as a function of scalar field $\phi$. Inserting the additional equation (\ref{s-equation}) in the autonomous system (\ref{autonomousGen}), the interaction term of model 1 (from equation (\ref{Interaction1})) then gives the following autonomous system

%%%%%%%%%%%%%%%%%%%%%%%%%%%%%%%%%%%%%%%%%%%%%%%%%%%%%%%%%%%%%%%%%%%%%%
\begin{eqnarray}
\begin{split}
   \frac{dx}{dN}& = \sqrt{\frac{3}{2}}ys - 3x + \frac{3}{2}x^{3} \frac{(1+z)}{(1-z)} (2-\omega_{m})+ \frac{3}{2}x\omega_{m}(1-y-z) \frac{(1+z)}{(1-z)} -\sqrt{\frac{3}{2}}\gamma(1-x^{2}-y-z),& \\
   \frac{dy}{dN}& = -\sqrt{6} xys + 3y\frac{(1+z)}{(1-z)} \left[x^{2}(2-\omega_{m}) + \omega_{m} (1-y-z) \right],&\\
   \frac{dz}{dN} &= -3z \left[x^{2}(2-\omega_{m}) +\omega_{m} (1-y-z) \right],&\\
   \frac{ds}{dN} &= -\sqrt{6} x (\mu^{2}-s^{2}),&~~\label{auto-Hyp-Int-1}
\end{split}
\end{eqnarray}
%%%%%%%%%%%%%%%%%%%%%%%%%%%%%%%%%%%%%%%%%%%%%%%
where $f(s)=(\mu^{2}-s^{2}).$
Since, $s=-\mu \tanh(\mu\phi)$ which implies that $-\mu\leq s \leq \mu$
so, the phase space is bounded. The compact phase space is as follows:
\begin{equation}\label{phase-boundary-Hyp}
\Psi_{\cosh}: \left \{(x,y,z): 0\leq x^{2}+y+z\leq 1, |x|\leq 1, 0 \leq y \leq1, 0 \leq z \leq1, |s| \leq \mu \right \}
\end{equation}
%%%%%%%%%%%%%%%%%%%%%%%%%%%%%%%%%%%%%%%%%%%%%%%%
The system (\ref{auto-Hyp-Int-1}) for the interaction (\ref{Interaction1}) with the hyperbolic potential admits the following thirteen critical points which arising from the valid physical region(\ref{phase-boundary-Hyp}) shows 4D phase space.

\begin{itemize}
\item  I. Critical Points : $ B1,~ B2 = ( \pm1, 0,0,\mu  ) $
\item II. Critical Points : $ B3,~ B4 = ( \pm1, 0,0,-\mu  ) $

\item  III. Critical Points : $ B5,~ B6 = \left( -\frac{\sqrt{6}}{3}\frac{\gamma}{(2-\omega_{m})}, 0, 0, \pm\mu \right ) $
\item  IV. Critical Points : $ B7,~ B8= \left ( \frac{\sqrt{6}}{2}\frac{\omega_{m}}{\gamma}, 0, \frac{1}{2}\frac{2\gamma^{2}-3\omega_{m}^{2}+6\omega_{m}}{\gamma^{2}}, \pm\mu \right).$

\item  V. Critical Point : $ B9 = \left( \frac{\mu}{\sqrt{6}}, 1-\frac{\mu^{2}}{6}, 0, \mu \right) $
\item  VI. Critical Point : $ B10 = \left( -\frac{\mu}{\sqrt{6}}, 1-\frac{\mu^{2}}{6}, 0, -\mu \right) $

\item VII. Critical Point : $ B11= \left (\frac{\sqrt{6}}{2}\frac{\omega_{m}}{(\gamma+\mu)}, \frac{\gamma}{\gamma+\mu}+\frac{3}{2}\frac{\omega_{m}(2-\omega_{m})}{(\gamma+\mu)^{2}}, 0, \mu \right) $
\item  VIII. Critical Point : $ B12= \left (\frac{\sqrt{6}}{2}\frac{\omega_{m}}{(\gamma-\mu)}, \frac{\gamma}{\gamma-\mu}+\frac{3}{2}\frac{\omega_{m}(2-\omega_{m})}{(\gamma-\mu)^{2}}, 0, -\mu \right) $
\item  IX. Line of critical Points : $ B13= \left ( 0,~ 1-z,~ z,~ 0  \right).$
\end{itemize}
%%%%%%%%%%%%%%%%%%%%%%%%%%%%%%%%%%%%%%%%%%%%%%%%%%
The critical points and their corresponding physical parameters are presented in the table \ref{modelQ1HYP} and the eigenvalues of linearized system for this interaction model with hyperbolic
potential are displayed in the table \ref{modelQ1HYP-Eigen}.
%%%%%%%%%%%%%%%%%%%%%%%%%%%%%%%%%%%%%%%%%%%%%%%%%%%%%%%%%%%%%%%%%%%%
%TCIMACRO{\TeXButton{B}{\begin{table}[tbp] \centering}}%
%BeginExpansion
\begin{table}[tbp] \centering
%EndExpansion
\caption{The Critical Points and their corresponding physical parameters for the interaction model $ Q1=\gamma \dot{\phi}\rho_{m}$ for hyperbolic potential are presented.}%
\begin{tabular}
[c]{ccccccc}\hline\hline
\textbf{Critical Points} & $(\mathbf{x,y,z,s})$ &$\mathbf{\omega_{\phi}}$ &
 $\mathbf{\Omega_{m}}$ & $\mathbf\Omega_{\phi}$ &
\textbf{q}\\\hline
$B1,B2  $ & $( \pm1, 0,0,\mu  )$ & $1$ &
$0$ & $1$ & $2$\\
$B3, B4  $ & $( \pm1, 0,0,-\mu  ) $ & $ 1  $ &
 $0 $ & $ 1 $& $2$ \\
$B5, B6  $ & $ \left( -\frac{\sqrt{6}}{3}\frac{\gamma}{(2-\omega_{m})},0,0,\pm\mu \right ) $ & $ 1  $  &
 $ \frac{12-12\omega_{m}+3\omega_{m}^{2}-2\gamma^{2}}{3(2-\omega_{m})^{2}} $  &       $ \frac{2\gamma^{2}}{3(2-\omega_{m})^{2}} $ &      $ \frac{2\gamma^{2}+8\omega_{m}-3\omega_{m}^{2}-4}{2(2-\omega_{m})} $ \\
 $B7, B8  $ & $ \left ( \frac{\sqrt{6}}{2}\frac{\omega_{m}}{\gamma} ,0, \frac{1}{2}\frac{2\gamma^{2}-3\omega_{m}^{2}+6\omega_{m}}{\gamma^{2}}, \pm\mu \right) $ & $ 1 $ &
 $ -\frac{3\omega_{m}}{\gamma^{2}} $ & $ \frac{3\omega_{m}^{2}}{2\gamma^{2}} $ & $ -1$ \\

 $B9, B10  $ & $ \left( \pm\frac{\mu}{\sqrt{6}}, 1-\frac{\mu^{2}}{6},0,\pm\mu \right)  $ & $ \frac{\mu^{2}}{3}-1 $ &
 $ 0 $ & $1 $ & $ -1+\frac{\mu^{2}}{2}$\\
 $B11  $ & $ \left (\frac{\sqrt{6}}{2}\frac{\omega_{m}}{(\gamma+\mu)}, \frac{\gamma}{\gamma+\mu}+\frac{3}{2}\frac{\omega_{m}(2-\omega_{m})}{(\gamma+\mu)^{2}},0 ,\mu \right) $ & $ \frac{-3\omega_{m}^{2}+3\omega_{m}+\gamma^{2}+\gamma\mu}{3\omega_{m}+\gamma^{2}+\gamma\mu} $ &
 $ \frac{\gamma\mu+\mu^{2}-3\omega_{m}}{(\gamma+\mu)^{2}} $ & $ \frac{\gamma\mu+\gamma^{2}+3\omega_{m}}{(\gamma+\mu)^{2}} $ & $ -\frac{2\gamma+2\mu-3\mu\omega_{m}}{2(\gamma+\mu)}$\\
 $B12  $ & $ \left (\frac{\sqrt{6}}{2}\frac{\omega_{m}}{(\gamma-\mu)}, \frac{\gamma}{\gamma-\mu}+\frac{3}{2}\frac{\omega_{m}(2-\omega_{m})}{(\gamma-\mu)^{2}},0 ,-\mu \right)$ & $  \frac{-3\omega_{m}^{2}+3\omega_{m}+\gamma^{2}-\gamma\mu}{3\omega_{m}+\gamma^{2}-\gamma\mu} $ & $ \frac{\mu^{2}-3\omega_{m}-\gamma\mu}{(\gamma-\mu)^{2}} $ &
 $ \frac{\gamma^{2}+3\omega_{m}-\gamma\mu}{(\gamma-\mu)^{2}} $ & $ -\frac{2\gamma-2\mu+3\mu\omega_{m}}{2(\gamma-\mu)}$ \\
 $B13  $ & $ \left ( 0, 1-z, z, 0  \right) $ & $ -1 $ &
 $ 0 $ & $ 1-z $ & $ -1$
 \\\hline\hline
\end{tabular}
\label{modelQ1HYP}
%TCIMACRO{\TeXButton{E}{\end{table}}}%
%BeginExpansion
\end{table}%
%EndExpansion
%%%%%%%%%%%%%%%%%%%%%%%%%%%%%%%%%%%%%%%%%%%%%
%TCIMACRO{\TeXButton{B}{\begin{table}[tbp] \centering}}%
%BeginExpansion
\begin{table}[tbp] \centering
%EndExpansion
\caption{The eigenvalues of the linearized system  for the interaction model (\ref{Interaction1}) for hyperbolic potential, where $m_{B11}=\sqrt{(36\mu^{2}-180\mu^{2}\omega_{m}+72\gamma\mu+81\mu^{2}\omega_{m}^{2}-36\gamma\mu\omega_{m}+36\gamma^{2}
-48\gamma^{3}\mu+432\omega_{m}^{2}-216\omega_{m}^{3}+144\omega_{m}\gamma^{2}-96\gamma^{2}\mu^{2}
-48\mu^{3}\gamma+72\omega_{m}^{2}\gamma\mu)}$,~~and~~\\
 $m_{B12}= \\
\sqrt{(36\mu^{2}-180\mu^{2}\omega_{m}-72\gamma\mu+81\mu^{2}\omega_{m}^{2}+36\gamma\mu\omega_{m}+36\gamma^{2}
+48\gamma^{3}\mu-216\omega_{m}^{3}+144\omega_{m}\gamma^{2}-96\gamma^{2}\mu^{2}+48\mu^{3}\gamma
-72\omega_{m}^{2}\gamma\mu+432\omega_{m}^{2})}$}%
\begin{tabular}
[c]{ccccccc}\hline\hline
\textbf{Critical Points} & $\mathbf{\lambda_{1}}$ &
 $\mathbf{\lambda_{2}}$ & $\mathbf{\lambda_{3}}$ & $\mathbf{\lambda_{4}}$ \\\hline
$B1  $ & $ -6 $ &  $ 6-3\omega_{m}+\gamma \sqrt{6}$ & $ 6-\mu \sqrt{6}$ & $2\sqrt{6}\mu$ \\
$B2  $ & $ -6 $ &  $ 6-3\omega_{m}-\gamma \sqrt{6}$ & $ 6+\mu \sqrt{6}$ & $-2\sqrt{6}\mu$ \\
$B3  $ & $ -6 $ &  $ 6-3\omega_{m}+\gamma \sqrt{6}$ & $ 6+\mu \sqrt{6}$ & $-2\sqrt{6}\mu$ \\
$B4  $ & $ -6 $ &  $ 6-3\omega_{m}-\gamma \sqrt{6}$ & $ 6-\mu \sqrt{6}$ & $2\sqrt{6}\mu$ \\
$B5  $ & $ \frac{2\gamma^{2}-3\omega_{m}^{2}+12\omega_{m}-12}{2(2-\omega_{m})} $ &  $ \frac{2\gamma^{2}-3\omega_{m}^{2}+6\omega_{m}+2\mu\gamma}{(2-\omega_{m})} $ & $ \frac{3\omega_{m}^{2}-2\gamma^{2}-6\omega_{m}}{(2-\omega_{m})}$& $-\frac{4\gamma\mu}{2-\omega_{m}}$ \\
$B6  $ & $ \frac{2\gamma^{2}-3\omega_{m}^{2}+12\omega_{m}-12}{2(2-\omega_{m})} $ &  $ \frac{2\gamma^{2}-3\omega_{m}^{2}+6\omega_{m}-2\mu\gamma}{(2-\omega_{m})} $ & $ \frac{3\omega_{m}^{2}-2\gamma^{2}-6\omega_{m}}{(2-\omega_{m})}$& $\frac{4\gamma\mu}{2-\omega_{m}}$ \\

$B7, B8  $ & $  \pm \frac{6\omega_{m}\mu}{\gamma } $ &
 $ \mp 3\frac{\omega_{m}\mu}{\gamma} $ & $ -\frac{3}{2}\frac{\gamma^{2}-\sqrt{\gamma^{4}-4\omega_{m}\gamma^{4}-12\omega_{m}^{2}\gamma^{2}+6\omega_{m}^{3} \gamma^{2}}}{\gamma^{2}}$ & $ -\frac{3}{2}\frac{\gamma^{2}+\sqrt{\gamma^{4}-4\omega_{m}\gamma^{4}-12\omega_{m}^{2}\gamma^{2}+6\omega_{m}^{3} \gamma^{2}}}{\gamma^{2}}  $ \\
$B9, B10  $ & $ -\mu^{2}$ & $  2\mu^{2} $ &
 $ -3+\frac{1}{2}\mu^{2}$& $ \pm\gamma\mu+\mu^{2}-3\omega_{m} $  \\
 $B11 $ & $ -\frac{3\omega_{m}\mu}{(\gamma+\mu)}$ & $ \frac{6 \omega_{m} \mu}{(\gamma+\mu)} $ &
 $ -\frac{3}{2}+\frac{3\mu\omega_{m}}{4(\gamma+\mu)}+\frac{m_{B11}}{4(\gamma+\mu)} $& $ -\frac{3}{2}+\frac{3\mu\omega_{m}}{4(\gamma+\mu)}-\frac{m_{B11}}{4(\gamma+\mu)}$ \\
$ B12 $ & $ \frac{3\omega_{m}\mu}{(\gamma-\mu)}$ & $ \frac{-6 \omega_{m} \mu}{(\gamma-\mu)} $ &
 $ -\frac{3}{2}-\frac{3\mu\omega_{m}}{4(\gamma-\mu)}+\frac{m_{B12}}{4(\gamma-\mu)} $& $ -\frac{3}{2}-\frac{3\mu\omega_{m}}{4(\gamma-\mu)}-\frac{m_{B12}}{4(\gamma-\mu)}$ \\
 $B13  $ & $ 0  $ & $ -3\omega_{m} $ &
 $ -\frac{3}{2}+\frac{1}{2}\sqrt{9-12\mu^{2}+12\mu^{2}z} $ & $ -\frac{3}{2}-\frac{1}{2}\sqrt{9-12\mu^{2}+12\mu^{2}z} $
 \\\hline\hline
\end{tabular}
\label{modelQ1HYP-Eigen}
%TCIMACRO{\TeXButton{E}{\end{table}}}%
%BeginExpansion
\end{table}%
%EndExpansion
%

$\bullet $ The solutions namely, B1, B2, B3, and B4 dominated by the kinetic
energy of the scalar field will exist for all parameter values involved in physical region.
The points show a decelerating phase ($q=2$) (see table \ref{modelQ1HYP}).
From the linear stability analysis, we observe that they are the hyperbolic
type critical points and all are saddle-like (unstable) critical points since any
two of the eigenvalues are of opposite sign (see table \ref{modelQ1HYP-Eigen}).
The points represent the solutions of standard cosmology ($z=0$).

$\bullet $ The points B5, and B6 correspond to the scaling solutions with
combination of both the fluids DE and DM, where DE behaves as stiff fluid in nature.
These points exhibit the  similar behavior with A3. The points exist for the
following parameter restrictions $0\leq \omega_{m} <2$ and  $\frac{3 \omega_{m} -6}{\sqrt{6}}\leq \gamma \leq \frac{6-3 \omega_{m} }{\sqrt{6}}$. Although, an acceleration is possible near the points
for $0\leq \omega_{m} <\frac{2}{3}$ and $-\frac{\sqrt{3 \omega_{m} ^2-8 \omega_{m} +4}}{\sqrt{2}}<\gamma <\frac{\sqrt{3 \omega_{m} ^2-8 \omega_{m} +4}}{\sqrt{2}}$,
 the points B5 and B6 represent  unstable (saddle-like) solutions (table \ref{modelQ1HYP-Eigen}).

$\bullet $ The points B7 and B8 are solutions with 5D corrections
in RSII brane scenario. They will exist for either
(i) $\omega_{m}=0$; ($\gamma<0$ or $\gamma>0$), or (ii) $\omega_{m}<0$; ($\gamma\leq \sqrt{\frac{3}{2}}\omega_{m}$ or $\gamma\geq -\sqrt{\frac{3}{2}}\omega_{m}$).
The DE is stiff fluid in nature. The expansion of the universe is always accelerating near those points (since $q=-1$). The points (for $\omega_m<0$) describing early inflationary scenario are always unstable in nature in the phase space while they represent the Big bang singularity for $\omega_{m}\longrightarrow 0.$

$\bullet $ The scalar field (i.e., DE) dominated (where DM is absent) solutions
(B9 and B10) are associated with a standard 4D behavior. They will exist for
 $0\leq \omega_{m} \leq 2$, and $-\sqrt{6}\leq \mu \leq \sqrt{6}$.
 The DE associated with these points represents a perfect fluid with equation of state $\omega_{\phi}=\frac{\mu^{2}}{3}-1$. The solutions are hyperbolic type critical points.
 The expansion of the universe will be accelerated near these points for $\mu^{2}<2$
 (see table \ref{modelQ1HYP}). The points, from hyperbolic potential, are unstable (saddle-like)
 in nature since one of the eigenvalues is always positive.  Therefore, we can conclude that
 the points are realized in the model with hyperbolic potential have different stability
 criteria from that of the model with exponential one (the point A4).

 $\bullet $  The existence conditions of the point B11 are: \\

 $(1)~~\omega_{m} =0: $\\
 $(a)~\mu <0,~~\mbox{and}~~\gamma <-\mu$\\
 $(b)~\mu =0,~~\mbox{and}~~ (\gamma <0,~~\mbox{or}~~ \gamma >0)$\\
 $(c)~\mu >0,~~\mbox{and}~~ \gamma >-\mu.$\\

 $(2)~0<\omega_{m} <2:$\\
 $(a)~~\mu \leq -\sqrt{6},~~\mbox{and}~~ \gamma \leq \frac{1}{2} \left(-2 \mu -\sqrt{6} \omega_{m} \right)$\\
 $(b)~ -\sqrt{6}<\mu <0,~~\mbox{and}~~ \gamma \leq \frac{3 \omega -\mu ^2}{\mu }$\\
 $(c)~0<\mu \leq \sqrt{6},~~\mbox{and}~~ \gamma \geq \frac{3 \omega_{m} -\mu ^2}{\mu }$\\
 $(d)~\mu >\sqrt{6},~~\mbox{and}~~ \gamma \geq \frac{1}{2} \left(\sqrt{6} \omega_{m} -2 \mu \right).$\\

 $(3)~\omega_{m} =2:$\\
 $(a)~\mu <-\sqrt{6},~~\mbox{and}~~ \gamma \leq \frac{1}{2} \left(-2 \mu -2 \sqrt{6}\right)$\\
 $(b)~ -\sqrt{6}\leq \mu <0,~~\mbox{and}~~ \gamma \leq \frac{6-\mu ^2}{\mu }$\\
 $(c)~0<\mu \leq \sqrt{6},~~\mbox{and}~~ \gamma \geq \frac{6-\mu ^2}{\mu }$\\
 $(d)~\mu >\sqrt{6},~~\mbox{and}~~ \gamma \geq \frac{1}{2} \left(2 \sqrt{6}-2 \mu \right).$\\

The matter- scalar field scaling solution  B11 is always accelerating in the phase
space since the conditions of acceleration are also the conditions for existence.
The point B11 in the phase space is unstable (saddle) in nature.

$\bullet $ The point B12 is the solution with the combination of both the DE and DM components.
The existence conditions for the point are:\\
$(1)~~\omega_{m} =0: $\\
$(a)~\mu <0,~~\mbox{and}~~ \gamma >\mu$\\
$(b)~\mu =0,~~\mbox{and}~~ (\gamma <0,~~\mbox{or}~~ \gamma >0)$\\
$(c)~\mu >0,~~\mbox{and}~~ \gamma <\mu$\\

$(2)~0<\omega_{m} <2:$\\
$(a)~\mu \leq -\sqrt{6},~~\mbox{and}~~ \gamma \geq \frac{1}{2} \left(2 \mu +\sqrt{6} \omega_{m} \right)$\\
$(b)~-\sqrt{6}<\mu <0,~~\mbox{and}~~ \gamma \geq \frac{\mu ^2-3 \omega_{m} }{\mu }$\\
$(c)~0<\mu \leq \sqrt{6},~~\mbox{and}~~ \gamma \leq \frac{\mu ^2-3 \omega }{\mu }$\\
$(d)~\mu >\sqrt{6},~~\mbox{and}~~ \gamma \leq \frac{1}{2} \left(2 \mu -\sqrt{6} \omega_{m} \right)$\\

$(3)~\omega_{m} =2:$\\
$(a)~\mu <-\sqrt{6},~~\mbox{and}~~ \gamma \geq \frac{1}{2} \left(2 \mu +2 \sqrt{6}\right)$\\
$(b)~ -\sqrt{6}\leq \mu <0,~~\mbox{and}~~ \gamma \geq \frac{\mu ^2-6}{\mu }$\\
$(c)~0<\mu \leq \sqrt{6},~~\mbox{and}~~ \gamma \leq \frac{\mu ^2-6}{\mu }$\\
$(d)~\mu >\sqrt{6},~~\mbox{and}~~ \gamma \leq \frac{1}{2} \left(2 \mu -2 \sqrt{6}\right).$\\

There exists an accelerating phase near the point B12 because the conditions
of acceleration are also the conditions for existence of the point.
The point is unstable (saddle) in nature.

$\bullet $  The line of critical points ($y=1-z$), namely, B13 represents
the solution with 5D corrections. This solution exists always in the phase
space (with $z\in[0,~1)$) and is dominated by the potential energy of the
scalar field ($\rho_{T}=V(\phi)$). The point is completely DE dominated solution.
It is the de Sitter like solution of early evolution of the universe ($\omega_{\phi}=-1$)
which is always accelerating. Although, this is a non-hyperbolic type point,
it has a 3D stable manifold for $0\leq\omega_{m}\leq2$,
$-\sqrt{\frac{3}{4(1-z)}}\leq\mu<0$ and $0<\mu\leq\sqrt{\frac{3}{4(1-z)}}$ with $0\leq z<1$,
unstable otherwise. Note that,
for $z=0$ the solution B13 can describe the late time de Sitter solution. This type of solution
can also be found in our earlier work \cite{S.Kr.Biswas2015a} and in Ref. \cite{Escobar1}.

%%%%%%%%%%%%%%%%%%%%%%%%%%%%%%%%%%%%%%%%%%%%%%%%%%%%%%%%%%%%%%%%%%%%%%%%%%%%%%%%%%%%%%%%%%%%%%%%%%%%

\subsection{Interaction Model 2:$ Q=\delta \frac{\rho_{m}}{H}\dot{\phi}^{2}$ }\label{Interaction2}

Now, in this sub-section we shall discuss the interaction model
\begin{equation}\label{Interaction2}
  Q=\delta \frac{\rho_{m}}{H}\dot{\phi}^{2}
 \end{equation}
%%%%%%%%%%%%%%%%%%%%%%%%%%%%%%%%%%%%%%%%%%%%
with the exponential potential and the hyperbolic potential respectively for the scalar field.
First, we present an autonomous system of ordinary differential equations in the presence of
exponential potential in three dimension  as a parameter $s$ which, here, has a constant value.

\subsubsection{Exponential potential:$ V=V_{0} exp(-s\phi)$ }

For the interaction (\ref{Interaction2}), substituting  $s$=constant in  (\ref{autonomousGen}),
we will get the three dimensional autonomous system in the form:
%%%%%%%%%%%%%%%%%%%%%%%%%%%%%%%%%%%%%%%%%%%%%%%%%%%%%%%%%%%%%%%%%%
\begin{eqnarray}
\begin{split}
   \frac{dx}{dN}& = \sqrt{\frac{3}{2}}ys - 3x + \frac{3}{2}x^{3} \frac{(1+z)}{(1-z)} (2-\omega_{m})+ \frac{3}{2}x\omega_{m}(1-y-z) \frac{(1+z)}{(1-z)}- 3\delta x(1-x^{2}-y-z),& \\
   \frac{dy}{dN}& = -\sqrt{6} xys + 3y\frac{(1+z)}{(1-z)}\left[x^{2}(2-\omega_{m}) + \omega_{m} (1-y-z)\right],&\\
   \frac{dz}{dN} & = -3z\left[x^{2}(2-\omega_{m}) +\omega_{m} (1-y-z)\right].&~\label{auto-Exp-Int-2}
\end{split}
\end{eqnarray}
%%%%%%%%%%%%%%%%%%%%%%%%%%%%%%%%%%%%%%%%%%%%%%%%%%%%%%%%%%%%%%%%%
Now, we first find out the critical points and then analyze their stability
(by applying linear stability theory) accordingly. We present the critical points
and the corresponding physical parameters  in the table \ref{modelQ2EXP} and the
eigenvalues of the linearized matrix in table \ref{modelQ2EXP-Eigen}.
The following are the critical points for the system (\ref{auto-Exp-Int-2}):
\begin{itemize}
\item  I. Critical Points : $ C1, C2 = ( \pm1, 0,0 ) $
\item  II. Critical Point : $ C3 = \left( \frac{s}{\sqrt{6}} , 1-\frac{s^{2}}{6},0  \right) $
\item  III. Critical Points : $ C4, C5 = \left( \pm \frac{\sqrt{2}}{2}\sqrt{\frac{\omega_{m}}{\delta}},0 , \frac{2-\omega_{m}+2 \delta}{2\delta }\right ) $
\item  IV. Critical Point : $ C6=  ( 0, 0, 0). $
\end{itemize}
%%%%%%%%%%%%%%%%%%%%%%%%%%%%%%%%%%%%%%%%%%%%%%%%%
%TCIMACRO{\TeXButton{B}{\begin{table}[tbp] \centering}}%
%BeginExpansion
\begin{table}[tbp] \centering
%EndExpansion
\caption{The existence of Critical Points and their corresponding physical parameters for the interaction model $ Q2=\delta \frac{\rho_{m}}{H}\dot{\phi}^{2}$ for exponential potential.}%
\begin{tabular}
[c]{ccccccc}\hline\hline
\textbf{Critical Points} & $(\mathbf{x,y,z})$ &$\mathbf{\omega_{\phi}}$ &
 $\mathbf{\Omega_{m}}$ & $\mathbf\Omega_{\phi}$ &
\textbf{q}\\\hline
$C1,C2  $ & $( \pm1, 0,0 )$ & $1$ &
$0$ & $1$ & $2$\\
$C3  $ & $ \left( \frac{s}{\sqrt{6}} , 1-\frac{s^{2}}{6},0  \right) $ & $ \frac{1}{3}s^{2}-1  $ &
 $ 0 $ & $ 1 $ & $ -1+\frac{1}{2}s^{2} $ \\
$C4, C5  $ & $ \left( \pm \frac{\sqrt{2}}{2}\sqrt{\frac{\omega_{m}}{\delta}},0 , \frac{2-\omega_{m}+2 \delta}{2\delta }\right ) $ & $ 1  $ &
 $-\frac{1}{\delta}$ & $ \frac{\omega_{m}}{2\delta}$ & $ -1 $ \\

 $C6  $ & $ ( 0, 0, 0)  $ & $ \nexists $ &
 $ 1 $ & $ 0 $ & $ -1+\frac{3\omega_{m}}{2}$
 \\\hline\hline
\end{tabular}
\label{modelQ2EXP}
%TCIMACRO{\TeXButton{E}{\end{table}}}%
%BeginExpansion
\end{table}%
%EndExpansion
%
%%%%%%%%%%%%%%%%%%%%%%%%%%%%%%%%%%%%%%%%
For the stability analysis, we have to find out the eigenvalues of the first
order perturbed  matrix and so, we present the eigenvalues for this model with
exponential potential  in  tabular form (see table \ref{modelQ2EXP-Eigen}).

  %TCIMACRO{\TeXButton{B}{\begin{table}[tbp] \centering}}%
%BeginExpansion
\begin{table}[tbp] \centering
%EndExpansion
\caption{The eigenvalues of the linearized system  for the interaction model $ Q2=\delta \frac{\rho_{m}}{H}\dot{\phi}^{2}$ for exponential potential}%
\begin{tabular}
[c]{ccccccc}\hline\hline
\textbf{Critical Points} & $\mathbf{\lambda_{1}}$ &
 $\mathbf{\lambda_{2}}$ & $\mathbf{\lambda_{3}}$ \\\hline
$C1, C2  $ & $ -6 $ &  $ 6-3\omega_{m}+6\delta$ & $ 6\mp s \sqrt{6}$ \\

$C3  $ & $ -s^{2} $ & $ \frac{1}{2}s^{2}-3  $ &
 $ \delta s^{2}+s^{2}-3\omega_{m} $  \\
 $C4, C5 $ & $ \mp s\sqrt{\frac{3\omega_{m}}{\delta}} $ & $ \frac{3}{\delta}\sqrt{-\omega_{m}\delta(2-\omega_{m}+2\delta)}  $ &
 $ -\frac{3}{\delta}\sqrt{-\omega_{m}\delta(2-\omega_{m}+2\delta)} $ \\

 $C6  $ & $ 3\omega_{m} $ & $ -3\omega_{m} $ &
 $ -3\delta-3+\frac{3}{2}\omega_{m} $
 \\\hline\hline
\end{tabular}
\label{modelQ2EXP-Eigen}
%TCIMACRO{\TeXButton{E}{\end{table}}}%
%BeginExpansion
\end{table}%
%EndExpansion
%

$\bullet $ The DE dominated solutions, namely,  C1 and C2  exist for all values of parameters.
Deceleration is always on face  ($q=2$) (see table \ref{modelQ2EXP}) for these solutions.
The hyperbolic point C1 is conditionally stable   in the following parameter region
(see table \ref{modelQ2EXP-Eigen}) $s>\sqrt{6}$, and $\delta<\frac{\omega_{m}}{2}-1 $,
whereas the point C2 will be stable in the range $s<-\sqrt{6}$ and $\delta<\frac{\omega_{m}}{2}-1.$

$\bullet $ The DE dominated solution C3 will exist for $-\sqrt{6}\leq s\leq\sqrt{6}$.
The DE behaves as perfect fluid, but acceleration will occur for $ s^{2}<2 $
(see table \ref{modelQ2EXP}). The point represents a non-hyperbolic solution for
$s=0$, $0<\omega_{m} \leq 2$ (hyperbolic otherwise). The hyperbolic point C3 describes
the stable solution if the following conditions are fulfilled
$\delta \in \mathbb{R}:$ \\
and either $(a)~\left(-\sqrt{6}<s<0,~~ 0\leq \omega_{m} \leq 2,~~ \delta <\frac{3 \omega_{m} -s^{2}}{s^{2}}\right)$\\

or, $(b)~\left(0<s<\sqrt{6},~~ 0\leq \omega_{m} \leq 2,~~ \delta <\frac{3 \omega_{m} -s^{2}}{s^{2}}\right).$\\

\begin{figure}
\centering
\subfigure[]{%
\includegraphics[width=7cm,height=5cm]{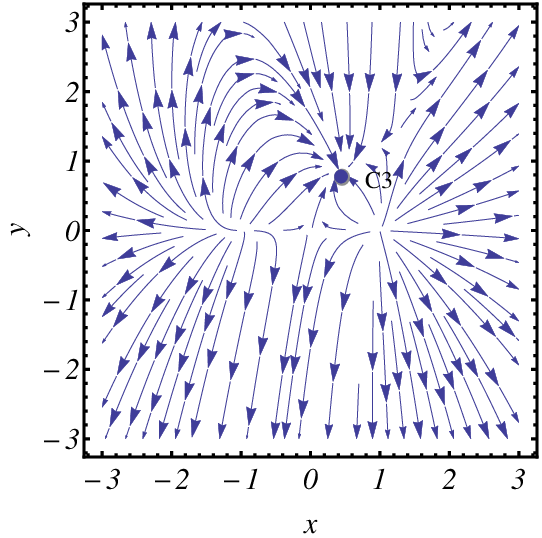}\label{fig:stable_C3_xy}}
\qquad
\subfigure[]{%
\includegraphics[width=7cm,height=5cm]{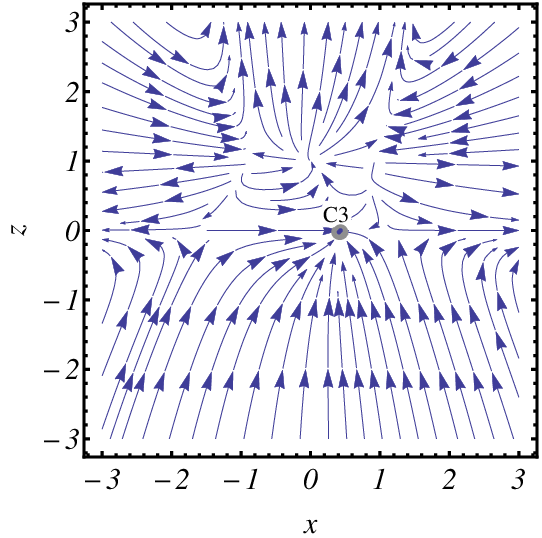}\label{fig:stable_C3_xz}}
\qquad
\subfigure[]{%
\includegraphics[width=7cm,height=5cm]{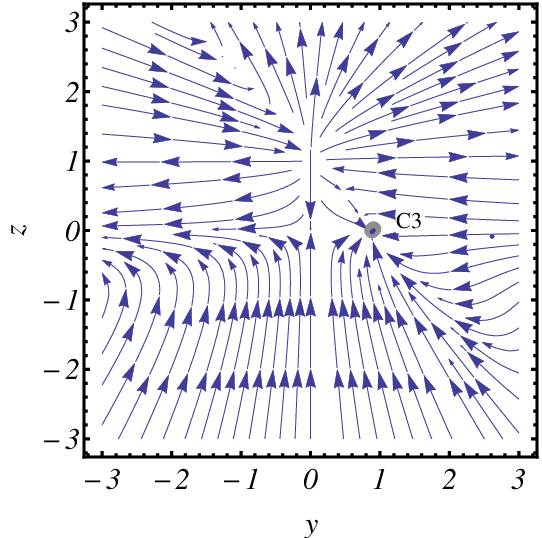}\label{fig:stable_C3_yz}}
\caption{The figure shows the phase portrait of the autonomous system (\ref{auto-Exp-Int-2}) for interaction (\ref{Interaction2}) with the parameters values $\delta =0.01$, $s=1$, and $\omega_{m} =1.01$. In panel (a), the point C3 exhibits a stable solution in the x-y phase plane. In (b), C3 shows its stability in the x-z plane, and  panel (c) shows that the point C3 is a stable solution in the y-z plane.}
\label{phasespace-coupling2-fig6-7-8}
\end{figure}

Although, the point represents an accelerated attractor in the above parameter space,
it does not solve the coincidence problem. The figure (\ref{phasespace-coupling2-fig6-7-8})
shows that the point C3 is stable attractor for parameter values $\delta =0.01$, $s=1$, $\omega =1.01.$ in different phase planes and the point C6 represents a unstable (saddle-like) solution.
%%%%%%%%%%%%%%%%%%%%%%%%%%%%%%%%%%%%%%%%%%%%%%%%%%%%%%%%%%%%%%%%%
%%%%%%%%%%%%%%%%%%%%%%%%%%%%%%%%%%%%%%%%%%%%%%%%%%%

$\bullet $ Critical points C4 and C5 will exist either for
(i) $\omega_{m}<0$, $\delta\leq \frac{\omega_{m}}{2}$, or for (ii) $\omega_{m}=0$, $\delta<0.$
These are the solutions with 5D corrections ($z\neq0$) in Brane scenario. Acceleration
is always possible around the points ($q=-1$). Here, DE is stiff fluid in nature.
The points are unstable(saddle) for $\omega_{m}<0$ (where matter mimics the phantom fluid) and
$\delta<0$, and describe the early inflationary transient state of the universe. On the
other hand, the critical points are non-hyperbolic in nature and become matter dominated in
phase space for $\omega_{m}\longrightarrow 0$.

$\bullet $ The point C6 represents completely DM dominated solution
($\Omega_{m}=1$) (see table \ref{modelQ2EXP}). We cannot conclude about the
nature of DE. Acceleration is possible for $\omega_{m}<\frac{2}{3}$ while the universe with a dust like matter (i.e., for $\omega_m=1$) is always decelerating ($q=\frac{1}{2}$).
This point shows an unstable (saddle-like) solution in the phase space of RS2 model,
since one of the eigenvalues is always positive (see table \ref{modelQ2EXP-Eigen}).

\subsubsection{Hyperbolic potential:$ V=V_{0} \cosh(\mu\phi)$ }

Now, considering  the scalar field related quantity $s$ as one of the dynamical variables,
the autonomous system for the interaction (\ref{Interaction2}) takes the form
%%%%%%%%%%%%%%%%%%%%%%%%%%%%%%%%%%%%%%%%%%%%%%%%%%%%%%%%%%%%%%%%%
\begin{eqnarray}
\begin{split}
   \frac{dx}{dN}& = \sqrt{\frac{3}{2}}ys - 3x + \frac{3}{2}x^{3} \frac{(1+z)}{(1-z)} (2-\omega_{m})+ \frac{3}{2}x\omega_{m}(1-y-z) \frac{(1+z)}{(1-z)} - 3\delta x(1-x^{2}-y-z),& \\
   \frac{dy}{dN}& = -\sqrt{6} xys + 3y\frac{(1+z)}{(1-z)}[x^{2}(2-\omega_{m}) + \omega_{m} (1-y-z)],& \\
   \frac{dz}{dN} &= -3z[x^{2}(2-\omega_{m}) +\omega_{m} (1-y-z)],& \\
   \frac{ds}{dN} &= -\sqrt{6} x (\mu^{2}-s^{2}),& \label{auto-Hyp-Int-2}
\end{split}
\end{eqnarray}
%%%%%%%%%%%%%%%%%%%%%%%%%%%%%%%%%%%%%%%%%%%%%%%%%%%%%%%%%%%%%%
where, phase space is described by the boundary (\ref{phase-boundary-Hyp}).
 The interaction  model 2 {\it i.e.}, $ Q2=\delta \frac{\rho_{m}}{H}\dot{\phi}^{2}$
 in the presence of the hyperbolic potential gives the  above four dimensional autonomous
 system of ordinary differential equations in which $s$ treated as a dynamical variable.
 The critical points for this system (\ref{auto-Hyp-Int-2}) are as follows:
\begin{itemize}
\item  I. Critical Points : $ D1, D2 = ( \pm1, 0,0,\mu  ) $
\item  II. Critical Points : $ D3, D4 = ( \pm1, 0,0,-\mu  ) $
\item  III. Critical Point : $ D5 = \left( \frac{\mu}{\sqrt{6}} , 1-\frac{\mu^{2}}{6},0,\mu \right) $
\item  IV. Critical Point : $ D6 = \left( -\frac{\mu}{\sqrt{6}} , 1-\frac{\mu^{2}}{6},0,-\mu \right) $
\item  V. Critical Points : $ D7, D8 = \left( \frac{1}{2}\sqrt{\frac{2\omega_{m}}{\delta}}, 0, \frac{2-\omega_{m}+2\delta}{2\delta}, \pm\mu \right ) $
\item  VI. Critical Points : $ D9, D10 = \left( -\frac{1}{2}\sqrt{\frac{2\omega_{m}}{\delta}}, 0, \frac{2-\omega_{m}+2\delta}{2\delta}, \pm\mu \right ) $

\item   VII. Curve of critical Point : $ D11 = ( 0,~0,~0,~s).$
\item  VIII. Line of critical Point : $  D12=  (0,~1-z,~z,~0). $
\end{itemize}
%%%%%%%%%%%%%%%%%%%%%%%%%%%%%%%%%%%%%%%%%%%%
The critical points and the corresponding physical parameters are displayed
in the table \ref{modelQ2HYP} and the eigenvalues for this interaction model with hyperbolic potential are
presented in the table \ref{modelQ2HYP-Eigen}.
%%%%%%%%%%%%%%%%%%%%%%%%%%%%%%%%%%%%%%%%%%%%%%%%%%%%%
%TCIMACRO{\TeXButton{B}{\begin{table}[tbp] \centering}}%
%BeginExpansion
\begin{table}[tbp] \centering
%EndExpansion
\caption{The Critical Points and their corresponding physical parameters for the interaction model $ Q2=\delta \frac{\rho_{m}}{H}\dot{\phi}^{2}$ for hyperbolic potential.}%
\begin{tabular}
[c]{ccccccc}\hline\hline
\textbf{Critical Points} & $(\mathbf{x,y,z,s})$ &$\mathbf{\omega_{\phi}}$ &
 $\mathbf{\Omega_{m}}$ & $\mathbf\Omega_{\phi}$ &
\textbf{q}\\\hline
$D1,D2  $ & $( \pm1, 0,0,\mu  )$ & $1$ &
$0$ & $1$ & $2$\\
$D3, D4  $ & $( \pm1, 0,0,-\mu  )  $ & $ 1  $ &
 $0 $ & $ 1 $& $2$ \\
 $D5, D6  $ & $ \left( \pm\frac{\mu}{\sqrt{6}} , 1-\frac{\mu^{2}}{6},0,\pm\mu \right) $ & $ \frac{\mu^{2}}{3}-1 $ &
 $ 0 $ & $1 $ & $ -1+\frac{\mu^{2}}{2}$\\
 $D7, D8  $ & $ \left( \frac{1}{2}\sqrt{\frac{2\omega_{m}}{\delta}}, 0, \frac{2-\omega_{m}+2\delta}{2\delta}, \pm\mu \right ) $ & $ 1 $ &
 $ -\frac{1}{\delta} $ & $ \frac{\omega_{m}}{2\delta} $ & $ -1$ \\
 $D9, D10  $ & $ \left( -\frac{1}{2}\sqrt{\frac{2\omega_{m}}{\delta}}, 0, \frac{2-\omega_{m}+2\delta}{2\delta}, \pm\mu \right ) $ & $ 1 $ &
 $ -\frac{1}{\delta} $ & $ \frac{\omega_{m}}{2\delta} $ & $ -1$ \\
 $D11  $ & $ ( 0,0,0,s) $ & $ \nexists $ &
 $ 1 $ & $ 0 $ & $ -1+\frac{3}{2}\omega_{m} $\\
 $D12  $ & $ (0,1-z,z,0) $ & $ -1 $ &
 $ 0 $ & $1-z $ & $ -1$\\

 \\\hline\hline
\end{tabular}
\label{modelQ2HYP}
%TCIMACRO{\TeXButton{E}{\end{table}}}%
%BeginExpansion
\end{table}%
%EndExpansion

%%%%%%%%%%%%%%%%%%%%%%%%%%%%%%%%%%%%%%%%%%

%TCIMACRO{\TeXButton{B}{\begin{table}[tbp] \centering}}%
%BeginExpansion
\begin{table}[tbp] \centering
%EndExpansion
\caption{The eigenvalues of the linearized system  for the interaction model $ Q1=\gamma \dot{\phi}\rho_{m}$ for hyperbolic potential. }%
\begin{tabular}
[c]{ccccccc}\hline\hline
\textbf{Critical Points} & $\mathbf{\lambda_{1}}$ &
 $\mathbf{\lambda_{2}}$ & $\mathbf{\lambda_{3}}$ & $\mathbf{\lambda_{4}}$ \\\hline
$D1, D2  $ & $ -6 $ &  $ 6-3\omega_{m}+6\delta$ & $ 6\mp\mu \sqrt{6}$ & $ \pm 2\sqrt{6}\mu$ \\
$D3, D4  $ & $ -6 $ &  $ 6-3\omega_{m}-6\delta$ & $ 6\pm \mu \sqrt{6}$ & $\mp 2\sqrt{6}\mu$ \\
$D5, D6  $ & $ -\mu^{2}$ & $  2\mu^{2} $ &
 $ -3+\frac{1}{2}\mu^{2}$& $ \mu^{2}\delta+\mu^{2}-3\omega_{m}$  \\

$D7, D8  $ & $ \frac{3}{\delta}\sqrt{-\omega_{m}\delta(2-\omega_{m}+2\delta)} $ &  $-\frac{3}{\delta}\sqrt{-\omega_{m}\delta(2-\omega_{m}+2\delta)} $ & $ \frac{\sqrt{3}\mu}{2\delta} \left(\pm \delta\sqrt{\frac{\omega_{m}}{\delta}}+3\sqrt{\omega_{m}\delta} \right)$ & $ -\frac{\sqrt{3}\mu}{2\delta} \left(\mp \delta\sqrt{\frac{\omega_{m}}{\delta}}+3\sqrt{\omega_{m}\delta} \right)$ \\

$D9, D10  $ & $ \frac{3}{\delta}\sqrt{-\omega_{m}\delta(2-\omega_{m}+2\delta)} $ &  $-\frac{3}{\delta}\sqrt{-\omega_{m}\delta(2-\omega_{m}+2\delta)} $ & $ \frac{\sqrt{3}\mu}{2\delta} \left(\mp \delta\sqrt{\frac{\omega_{m}}{\delta}}+3\sqrt{\omega_{m}\delta} \right)$ & $ -\frac{\sqrt{3}\mu}{2\delta} \left(\pm \delta\sqrt{\frac{\omega_{m}}{\delta}}+3\sqrt{\omega_{m}\delta} \right)$ \\

$D11 $ & $ 0 $ &  $ 3\omega_{m} $ & $ -3\omega_{m} $ & $-3\delta-3+\frac{3}{2}\omega_{m}$ \\
$D12  $ & $ 0  $ & $ -3\omega_{m} $ &
 $ -\frac{3}{2}+\frac{1}{2}\sqrt{9-12\mu^{2}+12\mu^{2}z} $ & $ -\frac{3}{2}-\frac{1}{2}\sqrt{9-12\mu^{2}+12\mu^{2}z} $
 \\\hline\hline
\end{tabular}
\label{modelQ2HYP-Eigen}
%TCIMACRO{\TeXButton{E}{\end{table}}}%
%BeginExpansion
\end{table}%
%EndExpansion
%

$\bullet $ The kinetic energy of the scalar field dominated (DM is absent)
solutions, namely, D1, D2, D3, and D4 always exist in the phase space.
These solutions represent a decelerating phase (since $q=2$). The DE behaves
as stiff fluid. The points are unstable (saddle-like) since one of eigenvalues
is always with opposite sign.

$\bullet $ Critical points D5 and D6 represent DE dominated solutions (where DM is absent)
which will exist for $\mu^{2}\leq6$. Here, DE is perfect fluid and acceleration is possible
for $\mu^{2}\leq2$. The points are unstable (saddle-like) in nature.

$\bullet $ The points D7, D8, D9, and D10 exist either for
(i) $\omega_{m}<0$, $\delta\leq \frac{\omega_{m}}{2}$, or for (ii) $\omega_{m}=0$, $\delta<0.$
The points will become matter dominated solutions for the condition $\omega_{m}=0$ (see table \ref{modelQ2HYP}).
In spite of that, there exists always an acceleration phase near those points. These are the solutions with
5D corrections approaching the early singularity of the universe where brane effects are prominent.
These points are non-hyperbolic in nature since all the eigenvalues are zero (for $\omega_{m}=0$) (see table \ref{modelQ2HYP-Eigen}). These points are unstable(saddle) in phase space
for $\omega_m<0$ and $\delta<0$ and provide an early inflationary
phase of the universe.

$\bullet $ The curve of critical points D11 represents always DM dominated solution
in space of phase in RSII scenario. We cannot conclude about the DE equation of state.
There exists an acceleration phase of universe near this point for $\omega_{m}<\frac{2}{3}$. For dust ($\omega_m=1$), the point decelerates ($q=\frac{1}{2}$).
The point is non-hyperbolic and saddle like in nature. So, D11 can describe the transient
phase of universe in the interacting DE context.

$\bullet $ Critical point D12 is a solution arising with 5D corrections in the
brane scenario. It exists always in the phase space (with $z\in[0,~1)$). The point
is completely DE dominated solution. It represents the de Sitter solution of early
evolution of universe ($\omega_{\phi}=-1$) and is always accelerating. This is non-hyperbolic
type point and has three dimensional stable subspace for $0\leq\omega_{m}\leq2$,
$-\sqrt{\frac{3}{4(1-z)}}\leq\mu<0$  and $0<\mu\leq\sqrt{\frac{3}{4(1-z)}}$ with $0\leq z<1$, otherwise it is unstable. The line of critical points can describe the late time de Sitter solutions dominated by DE for $z=0$.
%%%%%%%%%%%%%%%%%%%%%%%%%%%%%%%%%%%%%%%%%%%%%%%%%%%%%%%%%%%%%%%%%%

%%%%%%%%%%%%%%%%%%%%%%%%%%%%%%%%%%%%%%%%%%%%%%%%%%%%%%%%%%%%%%%%%%%%%%%%%%%%%%%%%%%%%%%%%%%%%%%%%%%%%%%%%%%%%%%%%%

\subsection{Interaction Model 3: Q3=$\sigma \frac{\rho_{m}^{2}}{H}$ }\label{Interaction3}

Now, we consider the interaction as
\begin{equation}\label{Interaction3}
Q=\sigma \frac{\rho_{m}^{2}}{H}
\end{equation}

\subsubsection{Exponential potential:$ V=V_{0} exp(-s\phi)$ }

Using the interaction term in equation (\ref{Interaction3}), exponential potential
of self interaction scalar field gives the autonomous system (\ref{autonomousGen}) as
%%%%%%%%%%%%%%%%%%%%%%%%%%%%%%%%%%%%%%%%%%%%%%%%%%%%%%%%%%%%%%
\begin{eqnarray}
\begin{split}
   \frac{dx}{dN}& = \sqrt{\frac{3}{2}}ys - 3x + \frac{3}{2}x^{3} \frac{(1+z)}{(1-z)} (2-\omega_{m})+ \frac{3}{2}x\omega_{m}(1-y-z) \frac{(1+z)}{(1-z)} - \frac{3}{2}\sigma \frac{(1-x^{2}-y-z)^{2}}{x}, & \\
   \frac{dy}{dN}& = -\sqrt{6} xys + 3y\frac{(1+z)}{(1-z)}\left[x^{2}(2-\omega_{m}) + \omega_{m} (1-y-z)\right],& \\
   \frac{dz}{dN} & =  -3z\left[x^{2}(2-\omega_{m}) +\omega_{m} (1-y-z)\right].&~\label{auto-Exp-Int-3}
\end{split}
\end{eqnarray}
%%%%%%%%%%%%%%%%%%%%%%%%%%%%%%%%%%%%%%%%%%%%%%%%%%%%%%%%%%%%%%%%%%%%%%%
To perform dynamical analysis, we first find out the critical points and
then analyze the linear stability theory accordingly. The critical
point for this system and the corresponding physical parameters are shown in the
table \ref{modelQ3EXP}. Using the interaction model (\ref{Interaction3}) with the exponential potential, the autonomous system (\ref{auto-Exp-Int-3}) produce the following critical points:
\begin{itemize}
\item  I. Critical Points : $ E1,~ E2 = ( \pm1,~ 0,~ 0 ) $
\item  II. Critical Point : $ E3 = \left( \frac{s}{\sqrt{6}}, 1-\frac{s^{2}}{6}, 0  \right) $
\item  III. Critical Points : $ E4,~ E5 = \left( \pm \sqrt{\frac{\sigma}{\omega_{m}-2+\sigma}} , 0, 0  \right) $
\item  IV. Critical Points : $ E6,~ E7 = \left( \pm \frac{\omega_{m}}{2}\sqrt{\frac{-2}{\sigma}}, 0 , \frac{2\sigma-2\omega_{m}+\omega_{m}^{2}}{2\sigma }\right ). $
\end{itemize}
%%%%%%%%%%%%%%%%%%%%%%%%%%%%%%%%%%%%%%%%%%%%%%%%%%%%%%%%%%%%%%%%%%%%
%TCIMACRO{\TeXButton{B}{\begin{table}[tbp] \centering}}%
%BeginExpansion
\begin{table}[tbp] \centering
%EndExpansion
\caption{The Critical Points and their corresponding physical parameters for the interaction model $  Q3=\sigma \frac{\rho_{m}^{2}}{H}$ for exponential potential.}%
\begin{tabular}
[c]{ccccccc}\hline\hline
\textbf{Critical Points} & $(\mathbf{x,y,z})$ &$\mathbf{\omega_{\phi}}$ &
 $\mathbf{\Omega_{m}}$ & $\mathbf\Omega_{\phi}$ &
\textbf{q}\\\hline
$E1,E2  $ & $( \pm1, 0,0 )$ & $1$ &
$0$ & $1$ & $2$\\
$E3  $ & $ \left( \frac{s}{\sqrt{6}} , 1-\frac{s^{2}}{6},0  \right) $ & $ \frac{1}{3}s^{2}-1  $ &
 $ 0 $ & $ 1 $ & $ -1+\frac{1}{2}s^{2} $ \\
$E4, E5  $ & $ \left( \pm \sqrt{\frac{\sigma}{\omega_{m}-2+\sigma}} , 0,0  \right)$ & $ 1  $ &
 $\frac{2-\omega_{m}}{2-\sigma-\omega_{m}}$ & $ \frac{\sigma}{-2+\sigma+\omega_{m}}$ & $ \frac{4+4\sigma-8\omega_{m}+3\omega_{m}^{2}}{2(-2+\sigma+\omega_{m})} $ \\

 $E6, E7  $ & $ \left( \pm \frac{\omega_{m}}{2}\sqrt{\frac{-2}{\sigma}},0 , \frac{2\sigma-2\omega_{m}+\omega_{m}^{2}}{2\sigma }\right ) $ & $ 1 $ &
 $ \frac{\omega_{m}}{\sigma} $ & $ -\frac{\omega_{m}^{2}}{\sigma} $ & $ -1 $
 \\\hline\hline
\end{tabular}
\label{modelQ3EXP}
%TCIMACRO{\TeXButton{E}{\end{table}}}%
%BeginExpansion
\end{table}%
%EndExpansion
%%%%%%%%%%%%%%%%%%%%%%%%%%%%%%%%%%%%%%%%%%%%%%%%%%%%%%%
For the stability analysis, we have to find out the eigenvalues of
linearized Jacobian matrix, and so we present the eigenvalues for this
model with exponential potential in the table \ref{modelQ3EXP-Eigen}.
%%%%%%%%%%%%%%%%%%%%%%%%%%%%%%%%%%%%%%%%%%%%%%%%%%%%%%%%%%%%%%%%%%%%%%%%%%
  %TCIMACRO{\TeXButton{B}{\begin{table}[tbp] \centering}}%
%BeginExpansion
\begin{table}[tbp] \centering
%EndExpansion
\caption{The eigenvalues of the linearized system  for the interaction model $ Q3=\sigma \frac{\rho_{m}^{2}}{H}$ for exponential potential}%
\begin{tabular}
[c]{ccccccc}\hline\hline
\textbf{Critical Points} & $\mathbf{\lambda_{1}}$ &
 $\mathbf{\lambda_{2}}$ & $\mathbf{\lambda_{3}}$ \\\hline
$E1,~ E2  $ & $ -6 $ &  $ 6-3\omega_{m}$ & $ 6\mp s \sqrt{6}$ \\

$E3  $ & $ -s^{2} $ & $ \frac{1}{2}s^{2}-3  $ &
 $ s^{2}-3\omega_{m} $  \\
 $E4,~ E5 $ & $ -6+3\omega_{m} $ & $  -\frac{3(2\sigma-2\omega_{m}+\omega_{m}^{2})}{-2+\sigma+\omega_{m}} $ &
 $ \frac{3(2\sigma-2\omega_{m}+\omega_{m}^{2})}{-2+\omega_{m}+\sigma} \mp s\sqrt{\frac{6\sigma}{-2+\omega_{m}+\sigma}}$ \\
$E6,~ E7 $ & $ \mp \omega_{m} s\sqrt{\frac{-3}{\sigma}} $ & $  -\frac{3}{2} (2-\omega_{m})+\frac{3}{2}\sqrt{(4-12\omega_{m}+\omega_{m}^{2})+\frac{4\omega_{m}^{2}}{\sigma}(2-\omega_{m})}$ &
 $ -\frac{3}{2} (2-\omega_{m})-\frac{3}{2}\sqrt{(4-12\omega_{m}+\omega_{m}^{2})+\frac{4\omega_{m}^{2}}{\sigma}(2-\omega_{m})} $ \\
 \\\hline\hline
\end{tabular}
\label{modelQ3EXP-Eigen}
%TCIMACRO{\TeXButton{E}{\end{table}}}%
%BeginExpansion
\end{table}%
%EndExpansion
%%%%%%%%%%%%%%%%%%%%%%%%%%%%%%%%%%%%%%%%%%%%%%%%%%%%%%%%%%%%%%%%%

$\bullet $ Completely DE dominated (DM is absent) solutions, namely,  E1 and E2
always exist in the phase space. Acceleration is not possible for those points.
They are unstable saddle like in nature.

$\bullet $ Critical point E3 exists for $s^{2}<6$. This point is completely dominated
by DE which behaves as perfect fluid. There exists an acceleration near this point
for $s^{2}<2$. The point is hyperbolic (for $s\neq 0$) in nature and will be stable
if the following conditions hold:
$s^{2}<min\{6,3\omega_{m}\}$, or in the range $0<\omega_{m}\leq 2$, and $-\sqrt{3\omega_{m}}<s<\sqrt{3\omega_{m}}$.
The figure (\ref{phasespace-coupling3-fig11}) shows that E3 is stable solution with the
parameter values $\sigma =-0.01$, $s=0.5$, $\omega_{m} =1.001.$
%%%%%%%%%%%%%%%%%%%%%%%%%%%%%%%%%%%%%%%%%%%%%%%%%%%%%%%%%%%%%%%%
\begin{figure}
\centering
\includegraphics[width=7cm,height=5cm]{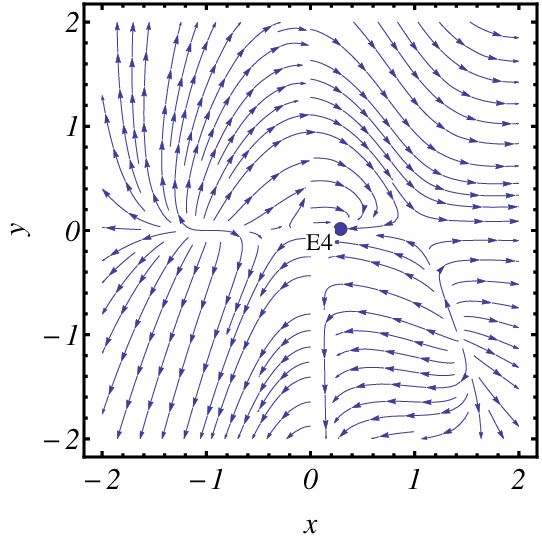}
\caption{In the phase plane of autonomous system (\ref{auto-Exp-Int-3}) with interaction (\ref{Interaction3}), the point E4 exhibits a stable solution for the parameters values  $\sigma =-0.1$, $s=3.5$, and $\omega_{m} =0.5 $.}
\label{phasespace-coupling3-fig9}
\end{figure}

\begin{figure}
\centering
\includegraphics[width=7cm,height=5cm]{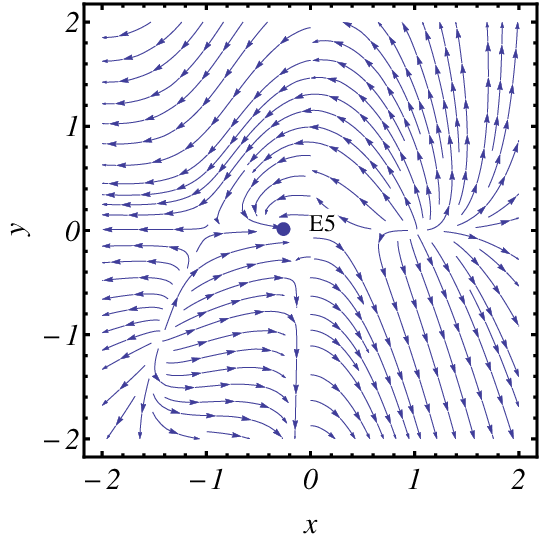}
\caption{Vector field shows the critical E5 of the autonomous system (\ref{auto-Exp-Int-3}) with interaction  (\ref{Interaction3}) is a stable solution for $\sigma =-0.1$,  $s=-3.5$, $\omega_{m} =0.5.$}
\label{phasespace-coupling3-fig10}
\end{figure}
%%%%%%%%%%%%%%%%%%%%%%%%%%%%%%%%%%%%%%%%%%%%%%%%%%%%%%%%%%%%%%%%%
\begin{figure}
\centering
\includegraphics[width=7cm,height=5cm]{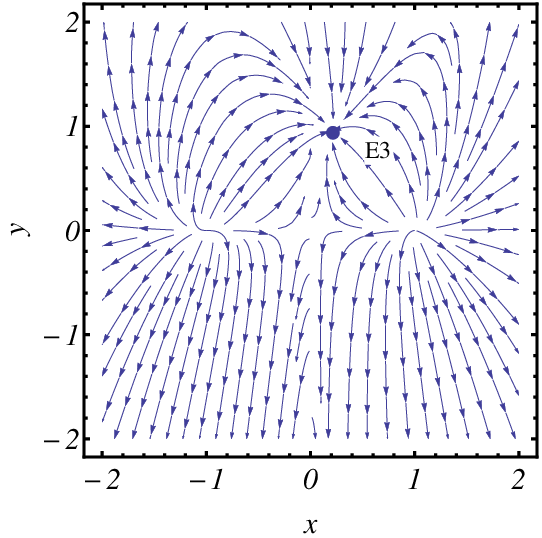}
\caption{The phase diagram of autonomous system (\ref{auto-Exp-Int-3}) with the interaction (\ref{Interaction3}) shows the point E3 is stable attractor for the parameters values $ \sigma =-0.01$, $s=0.5$, and  $\omega_{m} =1.001.$}
\label{phasespace-coupling3-fig11}
\end{figure}

%%%%%%%%%%%%%%%%%%%%%%%%%%%%%%%%%%%%%%%%%%%%%%%%%%%%%%%%%%%%%%%%%%%%%%
$\bullet $ Both the critical points E4 and E5 are in combination with the DE and DM components.
The DE behaves as stiff fluid. The points exist only when either
(a)~($0\leq \omega_{m} <2$,~ $\sigma \leq 0$),
or (b)~$\omega_{m} =2$,~~ ($\sigma <0$,~or ~ $\sigma >0$) hold. These are the DM dominated
solutions in uncoupled case, i.e., for $\sigma=0$. There will be an accelerating
phase of universe near the critical points if the following conditions are fulfilled
(a)~$\left( 0\leq \omega_{m} <\frac{2}{3},~~ \frac{1}{4} \left(-3 \omega_{m} ^2+8 \omega_{m} -4\right)<\sigma \leq 0\right )$
or, (b)~$\left( \omega_{m} =2,~ \sigma <0 \right), ~\mbox{or}~~(c)~\left( \omega_{m} =2,~~ \sigma >0 \right)$. Both the points are conditionally stable in some parameter region.
First, the point E4 will be stable in the parameter space:
$$0\leq \omega_{m} <\frac{2}{3},~~ \frac{1}{4} \left(-3 \omega_{m} ^2+8 \omega_{m} -4\right)<\sigma <0,~~ s>\sqrt{\frac{3}{2}} \sqrt{\frac{4 \sigma ^2+4 \sigma  \omega_{m} ^2-8 \sigma  \omega_{m} +\omega_{m} ^4-4 \omega_{m} ^3+4 \omega_{m} ^2}{\sigma  (\sigma +\omega_{m} -2)}},$$

and the point E5 is stable attractor for
 $$0\leq \omega_{m} <\frac{2}{3},~~ \frac{1}{4} \left(-3 \omega_{m} ^2+8 \omega_{m} -4\right)<\sigma <0,~~ s<-\sqrt{\frac{3}{2}} \sqrt{\frac{4 \sigma ^2+4 \sigma  \omega_{m} ^2-8 \sigma  \omega_{m} +\omega_{m} ^4-4 \omega_{m} ^3+4 \omega_{m} ^2}{\sigma  (\sigma +\omega_{m} -2)}}$$.\\
The figure (\ref{phasespace-coupling3-fig9}) shows E4 is stable attractor
for $ \sigma =-0.1$, $s=3.5$, $\omega_{m} =0.5.$
The point E5 is stable attractor for the state space $\sigma =-0.1$, $s=-3.5$, $\omega_{m} =0.5$
are shown in the figure (\ref{phasespace-coupling3-fig10}).

$\bullet $ Critical points E6 and E7 exist either for (i) $\omega_{m}=0$ and
$\sigma \in \{ \mathbb{R}- \{0\}\}$, or for (ii) $\omega_{m}<0$ and $\sigma\leq -\frac{\omega_{m}^{2}}{2}.$
This shows that the solutions are associated with 5D corrections in high energy regime
(where $\rho\gg \lambda$) describing the initial singularity as $z\rightarrow 1$ (for $\omega_{m}\longrightarrow 0$). The universe represented by these points is always accelerating in the RS model.
The points are non-hyperbolic (for $\omega_{m}=0$) and have a 2D stable manifold (see tables \ref{modelQ3EXP} and \ref{modelQ3EXP-Eigen}). On the other hand, for $\omega_{m}<0$, $\sigma<0$ the points are unstable in the phase space and represent early inflationary phase.
%%%%%%%%%%%%%%%%%%%%%%%%%%%%%%%%%%%%%%%%%%%%%%%%%%%%%%%%%%%%%%%%%%%%%%%%%%%%%%%%%%%%%%%%%%%%%

\subsubsection{Hyperbolic potential: $V=V_{0} \cosh(\mu\phi)$ }

For the hyperbolic potential and the interaction (\ref{Interaction3}),
the system (\ref{autonomousGen}) reduces to the form
%%%%%%%%%%%%%%%%%%%%%%%%%%%%%%%%%%%%%%%%%%%%%%%%%%%%%%%%%%%
\begin{eqnarray}
\begin{split}
   \frac{dx}{dN}& = \sqrt{\frac{3}{2}}ys - 3x + \frac{3}{2}x^{3} \frac{(1+z)}{(1-z)} (2-\omega_{m})+ \frac{3}{2}x\omega_{m}(1-y-z) \frac{(1+z)}{(1-z)}-\frac{3}{2}\sigma \frac{(1-x^{2}-y-z)^{2}}{x},& \\
   \frac{dy}{dN}& = -\sqrt{6} xys + 3y\frac{(1+z)}{(1-z)} \left[x^{2}(2-\omega_{m}) + \omega_{m} (1-y-z) \right],& \\
   \frac{dz}{dN} &= -3z \left[x^{2}(2-\omega_{m}) +\omega_{m} (1-y-z) \right],& \\
   \frac{ds}{dN} &= -\sqrt{6} x (\mu^{2}-s^{2}),&~~\label{auto-Hyp-Int-3}
\end{split}
\end{eqnarray}
%%%%%%%%%%%%%%%%%%%%%%%%%%%%%%%%%%%%%%%%%%%%%%%%%%%%%%%%%%%%
where the phase space  boundary satisfies (\ref{phase-boundary-Hyp}).
The critical points and the corresponding physical parameters for the
interaction model (\ref{Interaction3}) with the hyperbolic potential are shown in the table \ref{modelQ3HYP}.
The system (\ref{auto-Hyp-Int-3}) provides the following critical points:
\begin{itemize}
\item  I. Critical Points : $ F1,~ F2 = ( \pm1,~ 0,~ 0,~ \mu  ) $
\item  II. Critical Points : $ F3,~ F4 = ( \pm1,~ 0,~ 0,~ -\mu  ) $
\item  III. Critical Point : $ F5 = \left( \frac{\mu}{\sqrt{6}}, 1-\frac{\mu^{2}}{6},~ 0,~ \mu \right) $
\item  IV. Critical Point : $ F6 = \left( -\frac{\mu}{\sqrt{6}}, 1-\frac{\mu^{2}}{6},~ 0,~ -\mu \right) $
\item  V. Critical Points : $ F7,~ F8 = \left( \sqrt{\frac{\sigma}{\omega_{m}-2+\sigma}},~ 0,~ 0,~ \pm\mu \right ) $
\item  VI. Critical Points : $ F9,~ F10 = \left( -\sqrt{\frac{\sigma}{\omega_{m}-2+\sigma}},~ 0,~ 0,~ \pm\mu \right ) $
\item VII. Critical Points : $ F11,~ F12 = \left( \frac{\omega_{m}}{2}\sqrt{\frac{-2}{\sigma}}, 0,\frac{2\sigma+\omega_{m}^{2}-2\omega_{m}}{2\sigma}, \pm\mu \right ) $

\item  VIII. Critical Points : $ F13,~ F14 = \left( -\frac{\omega_{m}}{2}\sqrt{\frac{-2}{\sigma}}, 0,\frac{2\sigma+\omega_{m}^{2}-2\omega_{m}}{2\sigma}, \pm\mu \right ) $
\end{itemize}
%%%%%%%%%%%%%%%%%%%%%%%%%%%%%%%%%%%%%%%%%%%%%%%%%%%%%%%%%%%%%
%TCIMACRO{\TeXButton{B}{\begin{table}[tbp] \centering}}%
%BeginExpansion
\begin{table}[tbp] \centering
%EndExpansion
\caption{The existence of Critical Points and the corresponding physical parameters for the interaction model $ Q3=\sigma \frac{\rho_{m}^{2}}{H}$ for hyperbolic potential}%
\begin{tabular}
[c]{ccccccc}\hline\hline
\textbf{Critical Points} & $(\mathbf{x,y,z,s})$ &$\mathbf{\omega_{\phi}}$ &
 $\mathbf{\Omega_{m}}$ & $\mathbf\Omega_{\phi}$ &
\textbf{q}\\\hline
$F1,F2  $ & $( \pm1, 0,0,\mu  )$ & $1$ &
$0$ & $1$ & $2$\\
$F3, F4  $ & $( \pm1, 0,0,-\mu  ) $ & $ 1  $ &
 $0 $ & $ 1 $& $2$ \\
 $F5, F6  $ & $ \left(\pm \frac{\mu}{\sqrt{6}} , 1-\frac{\mu^{2}}{6},0,\pm\mu \right) $ & $ \frac{\mu^{2}}{3}-1 $ &
 $ 0 $ & $1 $ & $ -1+\frac{\mu^{2}}{2}$\\
 $F7, F8  $ & $ \left( \sqrt{\frac{\sigma}{\omega_{m}-2+\sigma}}, 0, 0 , \pm\mu \right ) $ & $ 1 $ &
 $ \frac{-2+\omega_{m}}{\omega_{m}-2+\sigma} $ & $ \frac{\sigma}{\omega_{m}-2+\sigma} $ & $ \frac{-8\omega_{m}+4+4\sigma+3\omega_{m}^{2}}{2(\omega_{m}-2+\sigma)}$ \\
 $F9, F10  $ & $ \left( -\sqrt{\frac{\sigma}{\omega_{m}-2+\sigma}}, 0, 0 , \pm\mu \right ) $ & $ 1 $ &
 $ \frac{-2+\omega_{m}}{\omega_{m}-2+\sigma} $ & $ \frac{\sigma}{\omega_{m}-2+\sigma} $ & $ \frac{-8\omega_{m}+4+4\sigma+3\omega_{m}^{2}}{2(\omega_{m}-2+\sigma)}$ \\
 $F11, F12  $ & $\left( \frac{\omega_{m}}{2}\sqrt{\frac{-2}{\sigma}}, 0,\frac{2\sigma+\omega_{m}^{2}-2\omega_{m}}{2\sigma} , \pm\mu \right )  $ & $ 1 $ &
 $ \frac{\omega_{m}}{\sigma} $ & $ -\frac{\omega_{m}^{2}}{2\sigma} $ & $ -1$ \\
 $F13, F14  $ & $ \left( -\frac{\omega_{m}}{2}\sqrt{\frac{-2}{\sigma}}, 0,\frac{2\sigma+\omega_{m}^{2}-2\omega_{m}}{2\sigma} , \pm\mu \right ) $ & $ 1 $ &
 $ \frac{\omega_{m}}{\sigma} $ & $ -\frac{\omega_{m}^{2}}{2\sigma} $ & $ -1$ \\
 \\\hline\hline
\end{tabular}
\label{modelQ3HYP}
%TCIMACRO{\TeXButton{E}{\end{table}}}%
%BeginExpansion
\end{table}%
%EndExpansion
%%%%%%%%%%%%%%%%%%%%%%%%%%%%%%%%%%%%%%%%%%%%%%%%
The eigenvalues for this interaction model with hyperbolic potential are
given in the table \ref{modelQ3HYP-Eigen}.

%TCIMACRO{\TeXButton{B}{\begin{table}[tbp] \centering}}%
%BeginExpansion
\begin{table}[tbp] \centering
%EndExpansion
\caption{The eigenvalues of the linearized system  for the interaction model $Q3=\sigma \frac{\rho_{m}^{2}}{H} $ for hyperbolic potential, where, $\sum_{\pm}= -\frac{3}{2} (2-\omega_{m}) \pm \frac{3}{2}\sqrt{(4-12\omega_{m}+\omega_{m}^{2})+\frac{4\omega_{m}^{2}}{\sigma}(2-\omega_{m})} $ }%
\begin{tabular}
[c]{ccccccc}\hline\hline
\textbf{Critical Points} & $\mathbf{\lambda_{1}}$ &
 $\mathbf{\lambda_{2}}$ & $\mathbf{\lambda_{3}}$ & $\mathbf{\lambda_{4}}$ \\\hline
$F1,~ F2  $ & $ -6 $ &  $ 6-3\omega_{m}$ & $ 6\mp\mu \sqrt{6}$ & $ \pm 2\sqrt{6}\mu$ \\
$F3,~ F4  $ & $ -6 $ &  $ 6-3\omega_{m}$ & $ 6\pm \mu \sqrt{6}$ & $\mp 2\sqrt{6}\mu$ \\
$F5,~ F6  $ & $ -\mu^{2}$ & $  2\mu^{2} $ &
 $ -3+\frac{1}{2}\mu^{2}$& $ \mu^{2}-3\omega_{m} $  \\

$F7,~ F8  $ & $ -3(2-\omega_{m}) $ &  $ \frac{3(2\sigma-2\omega_{m}+\omega_{m}^{2})}{2-\omega_{m}-\sigma} $ & $ \pm\frac{2 \mu \sqrt{6\sigma}}{\sqrt{\omega_{m}-2+\sigma}} $ & $ 6+\frac{3(\omega_{m}-2)^{2}}{\omega_{m}-2+\sigma}\mp \frac{\mu \sqrt{6\sigma}}{\sqrt{\omega_{m}-2+\sigma}}$ \\

$F9,~ F10  $ & $ -3(2-\omega_{m}) $ &  $ \frac{3(2\sigma-2\omega_{m}+\omega_{m}^{2})}{2-\omega_{m}-\sigma} $ & $ \mp \frac{2\mu \sqrt{6\sigma}}{\sqrt{\omega_{m}-2+\sigma}} $ & $ 6+\frac{3(\omega_{m}-2)^{2}}{\omega_{m}-2+\sigma} \pm \frac{\mu \sqrt{6\sigma}}{\sqrt{\omega_{m}-2+\sigma}}$ \\
$F11,~ F12 $ & $ \sum_{+} $ &  $ \sum_{-}  $ & $ - \frac{\sqrt{3} \mu\omega_{m}}{2\sqrt{-\sigma}} \left( 3 \mp \sqrt{-\sigma}\sqrt{\frac{1}{-\sigma}} \right) $ & $  \frac{\sqrt{3} \mu\omega_{m}}{2\sqrt{-\sigma}} \left( 3 \pm \sqrt{-\sigma}\sqrt{\frac{1}{-\sigma}} \right) $ \\
$F13,~ F14 $ & $ \sum_{+}  $ & $ \sum_{-} $ &
 $ - \frac{\sqrt{3} \mu\omega_{m}}{2\sqrt{-\sigma}} \left( 3 \pm \sqrt{-\sigma}\sqrt{\frac{1}{-\sigma}} \right) $ & $ \frac{\sqrt{3} \mu\omega_{m}}{2\sqrt{-\sigma}} \left( 3 \mp \sqrt{-\sigma}\sqrt{\frac{1}{-\sigma}} \right) $
 \\\hline\hline
\end{tabular}
\label{modelQ3HYP-Eigen}
%TCIMACRO{\TeXButton{E}{\end{table}}}%
%BeginExpansion
\end{table}%
%EndExpansion
%%%%%%%%%%%%%%%%%%%%%%%%%%%%%%%%%%%%%%%%%%%%%%%%%%%%%%%%%%%
There are fourteen critical points for the interaction model
(\ref{Interaction3}) with the hyperbolic potential of scalar field in the RSII model.

$\bullet $ F1, F2, F3, and F4 always exist in the phase space.
They are completely DE dominated solutions (DM is absent). Here,
DE behaves as stiff fluid. Accelerated expansion  is not possible around these points.
The points behave as unstable (saddle) solutions.

$\bullet $ The points F5 and F6 are completely DE dominated (DM absent) solutions,
and exist for $\mu^{2}\leq6$.
The DE is the perfect fluid. There exists an accelerating phase of the universe
near the points for $\mu^{2}\leq2$.
The points are saddle-like in nature since one of the eigenvalues is positive.

$\bullet $ The points F7, F8, F9, and F10 show the similar characteristics.
The existence criteria for these points are as follows:
$(a)~ 0\leq \omega_{m} <2,~~ \sigma \leq 0$
or$(b)~\omega_{m} =2,~~ (\sigma <0,~~\mbox{or}~~ \sigma >0)$. The points are the
solutions having both the DE and DM components. The DE is like stiff fluid in that case.
There exists an accelerating phase if the following conditions hold:
$(a)~\left(0\leq \omega_{m} <\frac{2}{3},~~ \frac{1}{4} \left(-3 \omega_{m} ^2+8 \omega_{m} -4\right)<\sigma \leq 0\right)$
or $(b)~ (\omega_{m} =2,~~ \sigma <0)$
or $(c)~(\omega_{m} =2,~~ \sigma >0)$. The points are unstable (saddle-like)
in nature in the physical region ( see table \ref{modelQ3HYP-Eigen}).

$\bullet $ F11, F12, F13, and F14 will have the similar nature as E6 and E7.

%%%%%%%%%%%%%%%%%%%%%%%%%%%%%%%%%%%%%%%%%%%%%%%%%%%%%%%%%%%%%%%%%%%%%%%%%%%%%%%%%%%%%%%

\section{Cosmological Implications}
\label{cosmological_implications}
%%%%%%%%%%%%%%%%%%%%%%%%%%%%%%%%%%%%%%%%%%%%%%%%%%%%%%%%%%%%
\begin{figure}
\centering
\subfigure[]{%
\includegraphics[width=9cm,height=7cm]{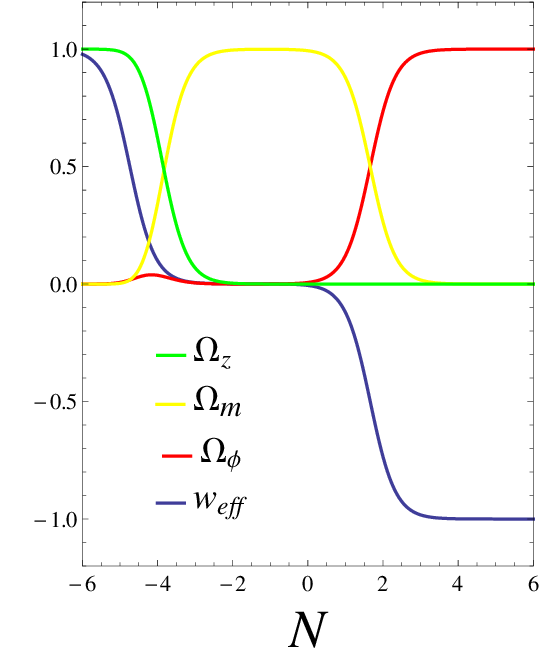}\label{fig:evolution_A}}
\caption{Time evolution of various physical quantities (the DE density parameter $\Omega_{\phi}$, the DM density parameter $\Omega_{m}$, the effective equation of state parameter $\omega_{eff}$, and the density of brane effects $\Omega_{z}(=z)$ versus $N=\ln a$) of the model (\ref{auto-Exp-Int-1}) for the interaction (\ref{Interaction1}) with parameter values $\omega_{m} =1.001$, $\gamma=0.001$, and $s=0.0005$ shows that the brane effects are prominent to the early universe, and after a long lasting matter domination the universe evolves towards an accelerated dark energy phase which phenomenologically agrees with observations.}
\label{phasespace-coupling1-evolution}
\end{figure}
%%%%%%%%%%%%%%%%%%%%%%%%%%%%%%%%%%%%%%%%%%%%%%%%%%%%%%%%%%%%

%%%%%%%%%%%%%%%%%%%%%%%%%%%%%%%%%%%%%%%%%%%%%%%%%%%%%%%%%%%%%%%%%%
As we have already discussed the phase space analysis of critical points
from the dynamics of RS2 brane scenario in the previous section.
Corresponding physical parameters and the stability of the
critical points are presented therein. We are now going to present
the cosmological implications of the points arising from the systems.
Cosmological evolutions of physically relevant quantities are shown
in the figure (\ref{phasespace-coupling1-evolution})
by choosing proper initial conditions. The figure confirms that the late time
attractors are the cosmological constant in the context of brane scenario for the parameter values
$\omega_{m} =1.001$, $\gamma=0.001$, and $s=0.0005$. It should be mentioned that
in the figure (\ref{phasespace-coupling1-evolution}), we use the symbol $\Omega_{z}$ for
$z$ (because, it is actually fractional contribution of brane tension to the total energy density)
as the contribution from brane towards the total energy density, where in figure, it can be observed that it
dominates in early universe and it has insignificant effect at late times.
Also the figure shows a clear trajectory describing an intermediate DM dominated
phase (where $\omega_{eff}=0$) which has transient nature in phase space to the late time DE domination ($\omega_{eff}=-1$). In this context, it should be noted that the critical points representing
the early inflation (for eg. point $B13$ with $z\longrightarrow 1$) are not the past attractor so
it is not possible to show graphically the variation of the cosmological parameters at the earlier inflationary scenario in fig. (\ref{phasespace-coupling1-evolution}). As the evolution of quantities
($\Omega_{\phi}$, $\Omega_{m}$, $\Omega_{z}$, and $\omega_{eff}$)
for all the models are similar, we can present only one model here.
In a cosmological perspective, the following scenarios are
realized by the properties of critical points:\\

{\bf a. Scalar field dominated era:}

The kinetic energy of scalar field dominated solutions, namely, A1 and A2
correspond to standard 4D behavior, where the expansion of universe
is always decelerated. Moreover, the solutions correspond to
conditional stable points in the phase space. On the other hand, the points
B1 $-$ B4 are always decelerating and saddle in nature. Similar scenarios can
also be obtained from the analysis of solutions C1, C2, D1 $-$ D4, E1, E2, and F1 $-$ F4
arising from different interactions with exponential as well as hyperbolic type potentials.
These are not interesting from physical point of view because of their unstable nature in phase space.

The solution represented by the point A4 corresponds to a scalar field dominated
era of the universe, where DE may be any kind of perfect fluid having equation of state
$\omega_{\phi}=\frac{s^{2}}{3}-1$. From the linear stability theory, we observe that the
point exhibits a stable solution for some parameter restrictions. It recovers
$\Lambda$CDM model of the universe for a constant
potential ({\it i.e.}, for $s=0$). The late time acceleration of the universe
near the point is realized for $s^{2}<2$, but it suffers from the coincidence
problem. Figures (\ref{phasespace-coupling1-fig1}) and (\ref{phasespace-coupling1-fig2})
explore that the point A4 is the stable attractor in the phase plane.
It should be mentioned that similar results are obtained by analyzing the points
C3 and E3 which all are achieved from exponential type potential. On the other hand,
for hyperbolic potential, the scalar field dominated solutions B9, B10, D5, D6, F5,
and F6 are physically insignificant as they are unstable points in the phase space for RS model.\\

{\bf b. Matter dominated era:}

A dark matter dominated ($\Omega_{m}=1$) era of the universe is described by the point C6.
There exists an accelerated universe if $\omega_{m}<\frac{2}{3}$, i.e., when
DM mimics as DE, otherwise it is decelerated.
From the linear stability analysis, we observe that the point C6 shows saddle-like nature,
since one of the eigenvalues is always positive. In this case, stability
in one eigen direction associated with negative eigenvalue while instability
will occur for positive eigenvalue. The point corresponds to a transient
stage of cosmic evolution since the universe can exit from this matter dominated era.
Thus, for a successful cosmological scenario, this phase should be relevant for
observed cosmic structure. Another point D11 will show the similar nature to that of
the point C6 in RSII model. Note that in an uncoupled case, i.e., when $\gamma\longrightarrow 0$,
the critical point A3 (in model 1) can give the matter dominated universe.\\

{\bf c. Matter-Scalar field scaling solutions:}

According to present observations our universe is currently undergoing an accelerated
expansion phase. This scenario of the universe can be realized in obtaining a critical point
representing the attractor solution which is accelerated with the similar order of dark energy
and dark matter $\Omega_{\phi}/\Omega_{m}=\mathcal{O}(1)$. Our study of interacting DE model
in brane dynamics reveals those points which are scaling attractor in phase space. Imposing some
restrictions on parameter, the point A5 behaves fully stable attractor which represents
the accelerating phase and the ratio of energy densities $r=\frac{\Omega_{\phi}}{\Omega_{m}}=
\frac{\gamma^{2}+\gamma s+3\omega_{m}}{s^{2}+\gamma s-3\omega_{m}}$ gives the possible
solution for coincidence problem. Thus, from the cosmological point of view,
the point A5 is relevant in the present context
by providing the possible explanation for the late time accelerated expansion of the universe.
Figures (\ref{phasespace-coupling1-fig3}), (\ref{phasespace-coupling1-fig4}),
and (\ref{phasespace-coupling1-fig5}) show that all trajectories enter into the point A5.
The similar results are also exhibited by the points A3, E4, and E5 for exponential potential,
where DE behaves as stiff fluid. The figures (\ref{phasespace-coupling3-fig9}) and (\ref{phasespace-coupling3-fig10})
exhibit the stable solutions E4 and E5 respectively.

On the other hand, for hyperbolic potential, matter-scalar field scaling
solutions represented by the points B11, B12, and F7 $-$ F10 also exist from
different interaction models. However, they are not
so much interesting  due to their unstable nature in the space of phase in RSII brane scenario.\\

%%%%%%%%%%%%%%%%%%%%%%%%%%%%%%%%%%%%%%%%%%%%%%%%%%%%%%%%%%%%%%%%%%%%%%%%%%%%%%%%%%%%%%%%%%%%%%%

The findings from the aforesaid discussions are as follows:
We have obtained several solutions in RSII brane scenario from which,
the solutions collectively can describe the overall evolution of the universe. Fist
of all, on some parameter restrictions, some of the points describe the big-bang singularity (for $\omega_{m}\longrightarrow 0$) as well as others represent early inflationary scenario of the universe (for $\omega_{m}<0$). Here, for both the cases, the universe is dominated by brane effects. From the study, early Big Bang singularity ($z\longrightarrow 1$) can be characterized by
the critical points A6, B7, B8, E6, E7, and F11 $-$ F14 which describe the ever accelerating
phase of the universe in the RS model. There exist some others solutions (C4, C5, and D7 $-$ D10) describing
the early accelerating model of universe with 5D corrections ($z\neq 0$).
 For all the cases, matter mimics the cosmological constant or the phantom fluid ($\omega_{m}\leq0$).
The points representing early inflationary scenario are unstable(saddle) in the phase space (for $\omega_{m}<0$)
and brane effects dominate the universe in the neighbourhood of all the points.

From the cosmological point of view, some points (B13 and D12, both for hyperbolic potential) are
describing early inflationary scenario of the universe, namely, the de Sitter solution
which gives early acceleration of universe. Also, these are the solutions with
5D corrections in the high energy regime. Here, the potential energy
dominates over the kinetic energy of the scalar field ($\rho_{T}=V$, $\rho_{m}=0$), so
the modified Friedmann equation takes the form
$3H^{2}=V(1+\frac{V}{2\lambda})$ and the high energy
expansion rate in RS model get enhanced relative to general relativistic rate
due to brane effect. Interestingly, it should be mentioned that the points can
describe the late time de Sitter solutions (DE dominated) for $z=0$ (without 5D corrections)
where brane effects dilute and GR is recovered.

Matter dominated solutions are interesting from the interacting DE point of view.
Note that a cosmological viable model should have a phase of matter domination
followed by an inflationary era of the universe. From our study,
we have obtained such type of solutions (C6 and D11)
describing an intermediate phase of the universe. These solutions are unstable
(saddle) and represent a phase transition.

Finally, the universe is undergoing through a late time accelerated phase which is
confirmed by several observations. We have also obtained some equilibrium points,
namely, A4, C3, and E3 representing
the late-time accelerated expansion of the universe, but they can not solve the
coincidence problem. In figure (\ref{phasespace-coupling2-fig6-7-8}), C3 is a future
attractor connecting to a matter dominated point C6 in its evolution.
Finally, some cosmologically interesting solutions, namely, matter-scalar field scaling solutions
have been obtained from our analysis. Among them, the point A5 is shown in
figures (\ref{phasespace-coupling1-fig3}), (\ref{phasespace-coupling1-fig4}),
and (\ref{phasespace-coupling1-fig5}) where all trajectories are entering into the
point A5 indicate the future attractor connecting
through unstable solution (past attractors). In particular, fig. (\ref{phasespace-coupling1-fig5}) shows
that A5 is a scaling solution satisfying $\Omega_{\phi}/\Omega_{m}=\mathcal{O}(1)$
with $\Omega_{\phi}=0.71$, $\Omega_{m}=0.29$, $\omega_{eff}=-0.53$ and it can solve the coincidence problem (Thus,
in the present cosmological context, A5 is relevant for interacting DE model, when DM behaves as dust).
In this context, the points A3, E4, and E5 are similar to that of A5.

 As we have observed in the model 1 with exponential potential that with
coupling $\gamma \longrightarrow 0$ the point $A3$ becomes DM dominated, and also there will have
the DE dominated solutions $A4$ and $A5$ followed by early inflationary solution $A6$. For
all the cases, the DM barotropic equation of state should be in the range $0\leq\omega_{m}\leq2$
and the universe will evolve from DM to DE state. We have obtained similar scenario in the model 3
where for uncoupled case, points $E4$, $E5$ are DM dominated and $E3$ is DE dominated solution
gives the late time acceleration of the universe. The points E6 and E7 represent the early inflation of
universe (for $\omega_{m}\longrightarrow 0$).
So, here the universe can also evolve from DM to DE state.
On the other hand, in the model 2 with the same exponential potential we have also
obtained a universe evolves from matter dominated era ($C6$) to DE dominated ($C3$) phase which has
been shown in the figure (\ref{phasespace-coupling2-fig6-7-8}) for $0\leq\omega_{m}\leq2$. Therefore,
in the parameter restriction of existence of critical points, some of the models have the nature
of evolution model from early inflation to late time DE domination connected through
a matter dominated era.\\

\begin{figure}
\centering
\subfigure[]{%
\includegraphics[width=7cm,height=5cm]{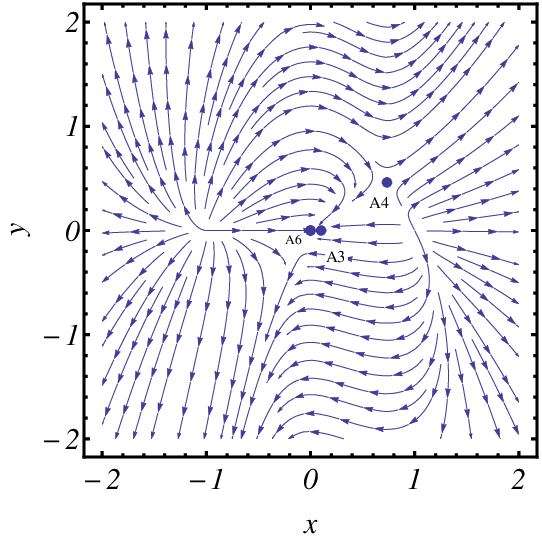}\label{fig:stable_A_unified}}
\qquad
\subfigure[]{%
\includegraphics[width=7cm,height=5cm]{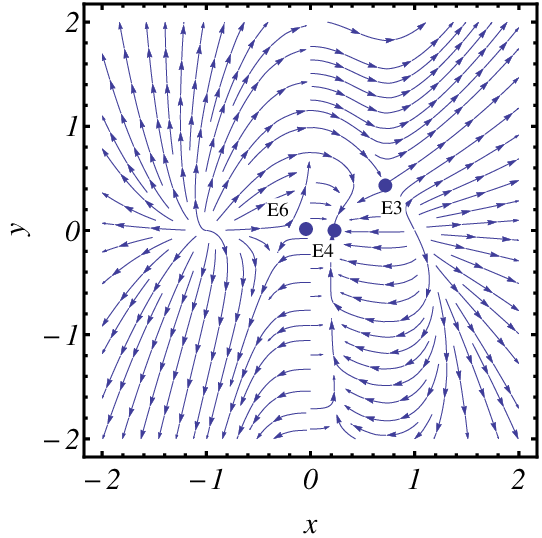}\label{fig:stable_E_unified}}
\caption{The figure shows the projection of phase portrait in x-y plane for the different systems. In panel (a), for the system (\ref{auto-Exp-Int-1}) with interaction (\ref{Interaction1}), the point A3 is late time stable solution,  A4 is the matter dominated saddle point, and A6 is unstable for $\omega_{m} =-0.001$, $\gamma=-0.1$, and $s=1.8$. In (b), for $\omega_{m} =-0.001$, $\sigma=-0.1$, and $s=1.8$, the point E4 is late time stable solution, E6, and E3 are early inflationary and matter dominated saddle point respectively for the system (\ref{auto-Exp-Int-3}) with interaction (\ref{Interaction3}).}
\label{phasespace-coupling1-3-unified}
\end{figure}

Moreover, the critical points presented in tables \ref{modelQ1EXP} and \ref{modelQ3EXP} corresponding
to two distinct DE models may describe the complete cosmic scenario, i.e., from inflationary
era to present late time acceleration.
In table \ref{modelQ1EXP}, the equilibrium points $A6$, $A4$ and $A3$ describe the early
inflationary era, the matter dominated epoch and the late accelerating phase of the universe
for $\omega_{m}$ to be negative but very close to zero. Also, in the limit $\omega_{m}\rightarrow 0$
the critical point $A6$ represents the big-bang singularity. Further, if one assumes
$\frac{|\omega_{m}|}{\gamma^{2}}$ to be constant and equal to $1/3$ then $A6$
can describe the late phase of evolution. Also, the free parameter $'s'$ in the critical
point $A4$ is restricted as $3<s^{2}<4$ to describe the matter dominated era. This can be shown numerically
in the figure (\ref{fig:stable_A_unified}) for the parameter values  $\omega_{m} =-0.001$, $\gamma=-0.1$, and $s=1.8$.
%%%%%%%%%%%%%%%%%%%%%%%%%%%%%%%%%%%%%%%%%%%
%%%%%%%%%%%%%%%%%%%%%%%%%%%%%%%%%%%%%%%%%%%%%
On the other hand, from table \ref{modelQ3EXP}, the whole evolution can be described
by the critical points as follows:
Inflationary era ($E6$, $E7$) $\longrightarrow$, Matter dominated era ($E3$)
$\longrightarrow$, Late time acceleration ($E4$, $E5$). However, the critical points $E6$, $E7$
exist provided both $\omega_{m}$ and $\sigma$ to be negative and $s^{2}$ will have the same restriction
(i.e, $3<s^{2}<4$) for the critical point $E3$ to represent the matter dominated era. This is
shown in the figure (\ref{fig:stable_E_unified}) for the parameter values $\omega_{m} =-0.001$, $\sigma=-0.1$, and $s=1.8$. Note that in both these models the perfect fluid behaves as DE while the DM
is described by the scalar field, i.e., the interacting fluids exchange their role as DM and DE.
Therefore, it can be claimed that the present choices of interactions are more physically viable
than those in Ref. \cite{S.Kr.Biswas2015a}, as it is possible to have a complete cosmic evolution for
the present models and hence, the present model can be considered as a generalization
of the previous work in Ref. \cite{S.Kr.Biswas2015a}.\\

We shall now examine whether the present brane world model with interacting two fluid
system may describe a complete cosmic scenario from initial inflationary era to the
present accelerated expansion of the universe. If we consider perfect fluid with constant
equation of state as dark matter interacting with dark energy, in the form of a minimally
coupled scalar field, then in table \ref{modelQ1HYP} the equilibrium point $B13$ corresponds
to early inflationary era for $z$ to be very close to unity (i.e., $z\longrightarrow 1$).
At that instant neither the DM nor the DE dominates, it is the extra brane term
(behaves as matter) dominates the evolution and here the values of $q$ and $\omega_{eff}$
are $-1$, i.e., the universe is in ever accelerating mode. In the same model,
the critical points $B5$ and $B6$ represent the matter dominated
era when the coupling parameter $\gamma$ to be very small. In this scenario, the DE
energy scalar field behaves as stiff fluid while the perfect fluid DM has the restricted
equation of state $2/3<\omega_{m}<2$. In particular, for the dust era of the model
$\omega_{m}$ is constrained as $2/3<\omega_{m}<1$ so that the DM behaves as normal fluid.
Both the critical points $B11$ and $B12$ describe the scaling solution of the model provided
$| \mu | \neq \gamma$. Lastly, the late time accelerating is described by the
equilibrium points $B9$, $B10$ provided $\mu^{2}<2$. The matter fully dominated by the DE
scalar field which behaves as an exotic matter for this choice of $\mu$. Further, from
the recent observational constraint \cite{Demianski} on the deceleration parameter (i.e., $q\simeq-0.5$)
the potential parameter $\mu$ can be restricted as $\mu^{2}\simeq 1$. It should be noted that
the critical point $B13$ with $z\longrightarrow 0$ may describe the present accelerated expansion
where the scalar field behaves as a cosmological constant and the perfect fluid has no effect
on the cosmological evolution but it is not of much interest as the value of $q (=-1)$ does
not match with observational data \cite{Demianski}.

%%%%%%%%%%%%%%%%%%%%%%%%%%%%%%%%%%%%%%%%%%%%%%%%%%%%%%%%%%%%%%%%%%%%%%%%%%%%%%%%%%%%%%%%%

\section{Summary and concluding remarks}
\label{summary}
In the previous sections, we have presented an exhaustive study of the
autonomous system describing the cosmic evolution in the background of brane
world gravity. We have taken three choices for the interaction term between the
two (dark) components of the cosmic substratum. The self interacting potential
function for the scalar field (describing dark energy component of the matter)
is chosen as exponential or hyperbolic in form. In general, the phase space for
the autonomous system is 4D, but it reduces to a 3D (\ref{phase-boundary-EXP})
phase space for the case of exponential type potential.

Several interesting cosmic scenarios can be obtained in the brane context, mainly,
at the early evolution of universe; on the other hand, DE cosmology is relevant
only for explaining the late time acceleration of the universe. Here, interacting
DE have been studied in the framework of brane cosmology to get a complete
evolutionary idea of universe from the dynamical systems perspective and
for that we have considered different interaction models from phenomenological ground.

We shall now summarize our main results for different interaction models with
the exponential and hyperbolic potential separately. We have observed that the
stability criteria of critical points arising from exponential potential are
essentially different from that of the hyperbolic type. The critical points,
achieved from every model for exponential potential, exhibits different characteristics in contrary
with the hyperbolic potential. For the exponential potential, the interactions provide solutions A1, A2 (from model (\ref{Interaction1})),
C1, C2 (from model (\ref{Interaction2})), and E1, E2 (from model (\ref{Interaction3}))
have the similar nature in the phase space. They correspond to the solutions, dominated
by the kinetic energy of the scalar field which are always decelerating in their cosmic evolution.
The points can describe stable solutions for some parameter restrictions.
On the other hand, for hyperbolic potential, the same interaction models provide the
critical points B1$-$B4; D1, D2; and F1$-$F4 dominated by kinetic energy,
exploring the decelerating phase of the universe are unstable (saddle)  in nature.

The scalar field dominated solutions namely, A4 (interaction (\ref{Interaction1})), C3 (interaction (\ref{Interaction2})), and E3((\ref{Interaction3})) with exponential potential exhibit
the same behavior from the cosmological point of view. For all the points, DE behaves
as perfect fluid having any equation of state $\omega_{\phi}=\frac{s^{2}}{3}-1$.
The expansion of universe will accelerate near these critical points for $s^{2}<2$ (decelerate otherwise).
The points represent the late time attractor for some parameter state but could not
alleviate the coincidence problem since $\mathcal{O}(\Omega_{\phi})=1$ in this case.
These are shown in the figures (\ref{phasespace-coupling1-fig1}), (\ref{phasespace-coupling2-fig6-7-8}), and (\ref{phasespace-coupling3-fig11}) respectively.

These type of critical points are also achieved from the hyperbolic potential.
The points B9, B10 (interaction model (\ref{Interaction1})); D5, D6 (Interaction (\ref{Interaction2}));
and F5, F6 (from model (\ref{Interaction3})) correspond to the solutions dominated by
the scalar field and are accelerating (for $\mu^{2}<2$) as the critical points
from exponential potential. But the points are not physically interested because of
their unstable (saddle-like) nature in the phase space ($x,~y,~z,~s$ in (\ref{phase-boundary-Hyp}))
where $'s'$ is treated as a dynamical variable.

The matter- scalar field scaling solutions A3 and A5 (from interaction (\ref{Interaction1}) );
E4 and E5 (from interaction model (\ref{Interaction3})) are realized from exponential potential.
 Depending on parameter restrictions, the points A3, A5,  E4, and E5
show the interesting behavior in the brane context. All behave as scaling late time attractors
which are accelerating (see the figures (\ref{phasespace-coupling1-fig3}), (\ref{phasespace-coupling1-fig4}), (\ref{phasespace-coupling1-fig5}), and (\ref{phasespace-coupling3-fig9}), (\ref{phasespace-coupling3-fig10})), and consequently together they can alleviate the coincidence problem.

The matter-scalar field scaling solutions are also achieved from the hyperbolic
potential but they do not show the same behavior with critical points as in the exponential potential.
The points provide the solutions which are never stable rather saddle in the physical
region ($x,~y,~z,~s$ in (\ref{phase-boundary-Hyp})).

The points with 5D corrections are produced for all cases of interactions and potentials.
They are non-hyperbolic in nature, linear stability theory fails to check their stability.
They represent the early big-bang model of universe.
Note that early inflationary de Sitter solutions can only be obtained from hyperbolic type potential.

Thus, we can conclude that the entire study is classified into two different aspects:
first is the study of interacting DE with three  different interaction terms in RSII brane
scenario choosing the self interaction potentials of scalar field  as exponential where
the quantity $s$ taken as constant, while the other is study with hyperbolic potential
where $s$ indicates  a dynamical variable (in the physical region (\ref{phase-boundary-Hyp})),
and hence the physical region would be 4D phase space. We have shown that for exponential
potential, the dynamics of RSII brane scenario produce some interesting critical points from
the cosmological point of view. The accelerated scaling attractors are realized for some
parameter restrictions in which the points could solve coincidence problem successfully.
However, in contrary to the exponential case, we have found that all the critical points
are unstable in nature in the four dimensional phase space.
Lastly, we can conclude that based on our study, the exponential potential
is more physically viable than hyperbolic potential from the point of view
of present cosmological scenario. Therefore, the present dynamical system analysis
shows that the present braneworld cosmological model describes a complete
cosmic picture from early inflationary era to present late time acceleration
and favours the recent observations.
%%%%%%%%%%%%%%%%%%%%%%%%%%%%%%%%%%%%%%%%%%%%%%%%%%%%%%%%%%%%%

\section*{Acknowledgments}

 The authors are thankful to IUCAA, Pune, India for providing research facilities as the work was done during a visit. One of the authors S.C is thankful to the UGC-DRS programme, in the department of Mathematics, Jadavpur University. S.K.B thanks Wompherdeiki Khyllep for helpful discussions on plotting figures.


\begin{thebibliography}{}

\bibitem{Akama1} K. Akama, An Early Proposal of 'Brane World', {\it Lect. Notes Phys.} \textbf{176} (1982), 267-271;

                 V. A. Rubakov, and M. E. Shaposhnikov, Extra Space-Time Dimensions: Towards a Solution to the Cosmological Constant Problem, {\it Phys. Lett. B} \textbf{125}  (1983), 139;
                 G. W. Gibbons, and D. L. Wiltshire, Space-Time as a Membrane in Higher Dimensions, {\it Nucl. Phys. B} \textbf{287} (1987), 717-742;
                 P. Horava, and E. Witten, Heterotic and type I string dynamics from eleven-dimensions, {\it Nucl. Phys. B} \textbf{460} (1996), 506-524;
                 P. Horava, and E. Witten, Eleven-dimensional supergravity on a manifold with boundary, {\it Nucl. Phys. B} \textbf{475} (1996), 94-114;
                 N. Arkani-Hamed, S. Dimopoulos, and G. R. Dvali, The Hierarchy problem and new dimensions at a millimeter, {\it Phys. Lett. B} \textbf{429} (1998), 263-272;
                 I. Antoniadis, N. Arkani-Hamed, S. Dimopoulos, and G. R. Dvali, New dimensions at a millimeter to a Fermi and superstrings at a TeV, {\it Phys. Lett. B} \textbf{436}  (1998), 257-263;
                 N. Kaloper, Bent domain walls as brane worlds, {\it Phys. Rev. D} \textbf{60} (1999), 123506.


\bibitem{Randall1} L. Randall, and R. Sundrum, A Large mass hierarchy from a small extra dimension, {\it Phys. Rev. Lett.} \textbf{83} (1999), 3370-3373.




\bibitem{Randall2}  L. Randall, and R. Sundrum, An Alternative to compactification, {\it Phys. Rev. Lett.} \textbf{83} (1999), 4690-4693.

\bibitem{Langlois1}  D Langlois, Brane cosmology: An Introduction, {\it Prog. Theor. Phys. Suppl.} \textbf{148} (2003), 181-212.
\bibitem{Garriga1}  J. Garriga and T. Tanaka, Gravity in the brane world, {\it Phys. Rev. Lett.} \textbf{84} (2000), 2778-2781;
                       S. B. Giddings, E. Katz, and L. Randall, Linearized gravity in brane backgrounds, {\it J. High Energy , Phys.} \textbf{ 0003} (2000), 023.
\bibitem{Riess1}  A. G. Riess  et al., New Hubble Space Telescope Discoveries of Type Ia Supernovae at $z>=1$: Narrowing Constraints on the Early Behavior of Dark Energy,  {\it Astrophys. J.} \textbf{659} (2007), 98-121.
\bibitem{Davis1}  T. M. Davis et al., Scrutinizing Exotic Cosmological Models Using ESSENCE Supernova Data Combined with Other Cosmological Probes,  {\it Astrophys. J.} \textbf{666} (2007), 716-725.
\bibitem{MichaelWood1}  W. Michael Wood-Vasey et al., Observational Constraints on the Nature of the Dark Energy: First Cosmological Results from the ESSENCE Supernova Survey, {\it Astrophys. J.} \textbf{666} (2007), 694-715.
\bibitem{Tegmark1}    M. Tegmark et al., Cosmological parameters from SDSS and WMAP, {\it Phys. Rev. D} \textbf{ 69 } (2004), 103501.
\bibitem{Jarosik1}   N. Jarosik et al., Seven-Year Wilkinson Microwave Anisotropy Probe (WMAP) Observations: Sky Maps, Systematic Errors, and Basic Results, {\it Astrophys. J. Suppl.} \textbf{192} (2011), 14.
\bibitem{Larson1}  D. Larson et al., Seven-Year Wilkinson Microwave Anisotropy Probe (WMAP) Observations: Power Spectra and WMAP-Derived Parameters,  {\it Astrophys. J. Suppl.} \textbf{192} (2011), 16.
\bibitem{Komatsu1}   E. Komatsu et al., Seven-Year Wilkinson Microwave Anisotropy Probe (WMAP) Observations: Cosmological Interpretation,  {\it Astrophys. J. Suppl.} \textbf{192} (2011), 18.
\bibitem{Sotiriou1}   T. P. Sotiriou and V. Faraoni, f(R) Theories Of Gravity, {\it Rev. Mod. Phys.} \textbf{82} (2010), 451-497;
                        S. Nojiri and S. D. Odintsov, Unified cosmic history in modified gravity: from F(R) theory to Lorentz non-invariant models, {\it Phys. Rept.} \textbf{505} (2011), 59-144;
                        Ya-Bo Wu et al., Thermodynamic laws for generalized f(R) gravity with curvature-matter coupling, {\it Phys. Lett. B} \textbf{ 717} (2012), 323-329;  S. Nojiri, S. D. Odintsov
                         and V. K. Oikonomou, Modified gravity theories on a nutshell: Inflation, bounce and late-time evolution, {\it Phys. Rept.} \textbf{692} (2017), 1-104 (arXiv:1705.11098 [gr-qc]).
\bibitem{Capozziello1}  S. Capozziello, Curvature quintessence, {\it Int. J. Mod. Phys. D}   \textbf{ 11} (2002), 483-492 (arXiv:gr-qc/0201033).

\bibitem{Nojiri1}       S. Nojiri and S. D. Odintsov, Modified f(R) gravity consistent with realistic cosmology: From matter dominated epoch to dark energy universe, {\it Phys. Rev. D}  \textbf{74} (2006), 086005.

\bibitem{Starobinsky1}  A. A. Starobinsky, A New Type of Isotropic Cosmological Models Without Singularity, {\it Phys. Lett. B}  \textbf{ 91} (1980), 99-102 ;
                        R. Kerner, Cosmology without singularity and nonlinear gravitational Lagrangians, {\it Gen. Relt. Grav.} \textbf{ 14}(1982) , 453-469;
                        J. D. Barrow and A. Ottewi, The Stability of General Relativistic Cosmological Theory, {\it J. Phys. A} \textbf{ 16} (1983), 2757;
                        V. Faraoni, Matter instability in modified gravity, {\it Phys. Rev. D} \textbf{ 74} (2006), 104017;
                        H. J. Schmidth, Fourth order gravity: Equations, history, and applications to cosmology, {\it Int. J. Geom. Meth. Mod. Phys.} \textbf{4} (2007), 209-248.
\bibitem{Nojiri2}       S. Nojiri and S. D. Odintsov, Modified gravity with $\ln R$ terms and cosmic acceleration {\it Gen. Rel. Grav.}  \textbf{ 36} (2004), 1765-1780; S. Nojiri and S. D. Odintsov, The Minimal curvature of the universe in modified gravity and conformal anomaly resolution of the instabilities,
                        {\it Mod. Phys. Lett. A} \textbf{ 19} (2004), 627-638;
                        M. C. B. Abdalla,  S. Nojiri and S. D. Odintsov, Consistent modified gravity: Dark energy, acceleration and the absence of cosmic doomsday, {\it Class. Quant. Grav.} \textbf{ 22} (2005), L35-L42.
\bibitem{Nojiri3}       S. Nojiri and S. D. Odintsov, Where new gravitational physics comes from: M Theory?, {\it Phys. Lett. B} \textbf{ 576} (2003), 5-11;
                        S. M. Carroll et al., Is cosmic speed - up due to new gravitational physics?, {\it Phys. Rev. D} \textbf{70} (2004), 043528;
                        S. Capozziello,  S. Nojiri, and S. D. Odintsov, Dark energy: The Equation of state description versus scalar-tensor or modified gravity, {\it Phys. Lett. B} \textbf{ 634} (2006), 93-100;
                        S. Nojiri, S. D. Odintsov and D. Saez-Gomes, Cosmological reconstruction of realistic modified F(R) gravities, {\it Phys. Lett. B} \textbf{ 681} (2009), 74-80.
\bibitem{Bengochea1}   G. R. Bengochea and R. Ferraro,  Dark torsion as the cosmic speed-up, {\it Phys. Rev. D} \textbf{ 79} (2005), 124019.


\bibitem{T. Padmanabhan2003} T. Padmanabhan, Cosmological constant: The Weight of the vacuum, {\it Phys. Rept.} \textbf{380} (2003), 235-320.
\bibitem{S. Weinberg1989} S. Weinberg, The Cosmological Constant Problem, {\it Rev. Mod. Phys.} \textbf{61} (1989), 1-23.
\bibitem{V. Sahni2000} V. Sahni and A. A. Starobinsky, The Case for a positive cosmological Lambda term {\it Int. J. Mod. Phys. D} \textbf{9} (2000),  373-444.


\bibitem{I. Zlatev1999} I. Zlatev, L.-M. Wang and P. J. Steinhardt, Quintessence, cosmic coincidence, and the cosmological constant, {\it Phys. Rev. Lett.} \textbf{82} (1999),  896-899
                       (arXiv: astro-ph/9807002).

\bibitem{C.G.Bohmer2008} C. G. Boehmer, G. Caldera-Cabral, R. Lazkoz and R. Maartens, Dynamics of dark energy with a coupling to dark matter, {\it Phys. Rev. D} \textbf{78} (2008), 023505.
\bibitem{C.Wetterich1995}  C. Wetterich, The Cosmon model for an asymptotically vanishing time dependent cosmological 'constant', {\it Astron. Astrophys.} \textbf{301} (1995), 321-328.
\bibitem{L.Amendola1999} L. Amendola, Scaling solutions in general nonminimal coupling theories, {\it Phys. Rev. D} \textbf{60} (1999), 043501.




\bibitem{Oliveras1} G. Olivares, F. Atrio- Barandela and D. Pavon, Observational constraints on interacting quintessence models, {\it Phys. Rev. D} \textbf{71} (2005), 063523.

\bibitem{Oliveras2} G. Oliveras, F. Atrio- Barandela and D. Pavon,
Matter density perturbations in interacting quintessence models, {\it Phys. Rev. D} \textbf{74} (2006), 043521.
\bibitem{Das1}  S. Das, P. S. Corasaniti and J. Khoury, Super-acceleration as Signature of Dark Sector Interaction, {\it Phys. Rev. D} \textbf{73} (2006), 083509.

\bibitem{Amendola1} L. Amendola, M. Gasperini and F. Piazza, SNLS data are consistent with acceleration at z=3, {\it Phys. Rev. D} \textbf{74} (2006), 127302.


\bibitem{Yuri.L.Bolotin2014} Y. L. Bolotin, A. Kostenko, O. A. Lemets and D  A. Yerokhin, Cosmological Evolution With Interaction Between Dark Energy And Dark Matter,  {\it Int. J. Mod. Phys. D} \textbf{24} (2015), no.03 1530007 (arXiv: 1310.0085 [astro-ph.CO]).
\bibitem{Andre A.Costa2014}  A. A. Costa, X. D. Xu, B. Wang, E. G. M. Ferreira and E. Abdalla, Testing the Interaction between Dark Energy and Dark Matter with Planck Data, {\it Phys. Rev. D} \textbf{89}, 103531 (2014), no.10, 103531.
\bibitem{M.Khurshudyan2015}  M. Khurshudyan and R. Myrzakulov, Phase space analysis of some interacting Chaplygin gas models, {\it Eur. Phys. J. C}  \textbf{77} (2017), no.2, 65 (arXiv:1509.02263 [gr-qc]).
\bibitem{S.Kr.Biswas2015a} 	S. Kr. Biswas and S. Chakraborty, Dynamical systems analysis of an interacting dark energy model in the brane scenario, {\it Gen. Rel. Grav.}   \textbf{47} (2015), 22.

\bibitem{S.Kr.Biswas2015b}  S. Kr. Biswas and S. Chakraborty, Interacting Dark Energy in f(T) cosmology : A Dynamical System analysis, {\it Int. J.  Mod. Phys. D.} \textbf{24} (2015), no.07, 1550046.




\bibitem{N.Tamanini2015} N. Tamanini, Phenomenological models of dark energy interacting with dark matter {\it Phys. Rev. D} \textbf{92} (2015), no.4, 043524 (arXiv: 1504.07397 [gr-qc]).

\bibitem{Xi-ming Chen2009} X. Chen, Y. Gong and E.N. Saridakis, Phase-space analysis of interacting phantom cosmology, {\it JCAP} \textbf{0904} (2009), 001 (arXiv:0812.1117[gr-qc]).
\bibitem{T.Harko2013} T. Harko and F. S. N. Lobo, Irreversible thermodynamic description of interacting dark energy-dark matter cosmological models, {\it Phys. Rev. D} \textbf{87} (2013), no.4, 044018 (arXiv:1210.3617[gr-qc]).


\bibitem{Nunes2016} R. C. Nunes, S. Pan and E. N. Saridakis, New constraints on interacting dark energy from cosmic chronometers, {\it Phys. Rev. D} \textbf{94} (2016), no.2, 023508  (arXiv:1605.01712 [astro-ph.CO]).
\bibitem{Wang2016} B. Wang, E. Abdalla, F. Atrio-Barandela and D. Pavon, Dark Matter and Dark Energy Interactions: Theoretical Challenges, Cosmological Implications and Observational Signatures, {\it Rept. Prog. Phys.} {\textbf 79} (2016), no.9, 096901 (arXiv:1603.08299 [astro-ph.CO]).
\bibitem{Landim12016} R. C. G. Landim, Dynamical analysis for a vector-like dark energy, {\it Eur. Phys. J. C} \textbf{76} (2016), no.9, 480 (arXiv:1605.03550 [hep-th]).
\bibitem{Landim22016} R. C. G. Landim,
Coupled dark energy: a dynamical analysis with complex scalar field, {\it Eur. Phys. J. C} \textbf{76} (2016), no.1, 31 (arXiv:1507.00902 [gr-qc]).




\bibitem{Maeda1} K. Maeda, Brane quintessence, {\it Phys. Rev. D} \textbf{64} (2001), 123525.
\bibitem{Pedro1} P. F. Gonzalez-Diaz, Quintessence in brane cosmology,  {\it Phys. Lett. B} \textbf{481} (2000), 353-359.
\bibitem{Majumdar1} A. S. Majumdar, From brane assisted inflation to quintessence through a single scalar field, {\it Phys. Rev. D} \textbf{64} (2001), 083503.
\bibitem{Nunes1} N. J. Nunes, and E. J. Copeland, Tracking quintessential inflation from brane worlds, {\it Phys. Rev. D} \textbf{66 } (2002), 043524.
\bibitem{Sami1} M. Sami and N. Dadhich, Unifying brane world inflation with quintessence, TSPU Bulletin 44N7 (2004) 25-36.
\bibitem{Gonzalez1} T. Gonzalez, T. Matos, I. Quiros and A.
                        V. Gonzalez, Self-interacting Scalar Field Trapped in a Randall-Sundrum Braneworld: The Dynamical Systems Perspective,  {\it Phys. Lett. B}  \textbf{676} (2009), 161-167.
\bibitem{Leyva1} Y. Leyva, D. Gonzalez, T. Gonzalez, T.
                        Matos and I. Quiros, Dynamics of a self-interacting scalar field trapped in the braneworld for a wide variety of self-interaction potentials, {\it Phys. Rev. D} \textbf{80} (2009), 044026.

\bibitem{Escobar1} D. Escobar, C. R. Fadragas, G. Leon and Y. Leyva, Phase space analysis of quintessence fields trapped in a Randall-Sundrum Braneworld: a refined study, {\it Class. Quant.  Grav.} \textbf{29} (2012), 175005 ( arXiv: 1110.1736v3 [gr-qc]).
\bibitem{Jibitesh1} J. Dutta and H. Zonunmawia, Complete cosmic scenario in the Randall-Sundrum braneworld from the dynamical systems perspective, {\it Eur. Phys. J. Plus} \textbf{130} (2015), no.11, 221 (arXiv:1601.00283 [gr-qc]).


\bibitem{Copeland1} E. J. Copeland, A. R. Liddle and D. Wands, Exponential potentials and cosmological scaling solutions, {\it Phys. Rev. D} \textbf{57} (1998), 4686-4690.
\bibitem{Aguirregabiria1} J. M. Aguirregabiria, and R. Lazkoz, Tracking solutions in tachyon cosmology {\it Phys. Rev. D} \textbf{69} (2004), 123502.
\bibitem{Lazkoz1} R. Lazkoz, G. Leon and I. Quiros, Quintom cosmologies with arbitrary potentials, {\it Phys. Lett. B} \textbf{649} (2007), 103-110.
\bibitem{Fang1} W. Fang, Y. Li, K. Zhang, and H. Qing Lu,  Exact Analysis of Scaling and Dominant Attractors Beyond the Exponential Potential, {\it Class. Quant. Grav.} \textbf{26 } (2009), 155005.
\bibitem{Leon1} G. Leon, On the Past Asymptotic Dynamics of Non-minimally Coupled Dark Energy, {\it Class. Quant. Grav.} \textbf{26} (2009), 035008.
\bibitem{Leon2} G. Leon, P. Silveira and C. R. Fadragas, Phase-space of flat Friedmann-Robertson-Walker models with both a scalar field coupled to matter and radiation,  arXiv: 1009.0689 [gr-qc].
\bibitem{Leon3} G. Leon and C. R. Fadragas, Cosmological dynamical systems, LAP Lambert Academic Publishing (2012), ( arXiv:1412.5701 [gr-qc]).




\bibitem{Campos1} A. Campos and C. F. Sopuerta, Evolution of cosmological models in the brane world scenario, {\it Phys. Rev. D} \textbf{63} (2001), 104012.
\bibitem{Campos2} A. Campos and C. F. Sopuerta, Bulk effects in the cosmological dynamics of brane world scenarios, {\it Phys. Rev. D} \textbf{64} (2001), 104011.
\bibitem{Coley1}  A. A. Coley, Dynamics of brane world cosmological models,  {\it Phys. Rev. D} \textbf{66} (2002), 023512.
\bibitem{Coley2}  A. A. Coley, The Dynamics of brane-world cosmological models, {\it Can. J. Phys.} \textbf{83} (2005) 475  (astro-ph\textbf{/0504226}).


\bibitem{Goheer1} N. Goheer and P. K. S. Dunsby,  Exponential potentials on the brane, {\it Phys. Rev. D} \textbf{67} (2003),  103513.
\bibitem{Goheer2} N. Goheer and P. K. S. Dunsby, Brane world dynamics of inflationary cosmologies with exponential potentials, {\it Phys. Rev. D} \textbf{66} (2002), 043527.



\bibitem{Odintsov2017} S. D. Odintsov and  V. K. Oikonomou, Autonomous dynamical system approach for
f(R) gravity, {\it Phys. Rev. D} \textbf{96} (2017), 104049  (arXiv:1711.02230 [gr-qc]).
\bibitem{Oikonomou2018} V. K. Oikonomou, Autonomous dynamical system approach for inflationary Gauss-Bonnet modified Gravity, {\it Int. J. Mod. Phys. D} \textbf{27} (2018), no. 05, 1850059  (arXiv:1711.03389 [gr-qc]).
\bibitem{Odintsov2018a} S. D. Odintsov and V. K. Oikonomou, Dynamical systems perspective of cosmological finite-time singularities in f(R) gravity and interacting multifluid cosmology, {\it Phys. Rev. D} \textbf{98} (2018), 024013  (arXiv:1806.07295 [gr-qc]).
\bibitem{Kleidis2018} K. Kleidis and V. K. Oikonomou, Autonomous dynamical system description of de Sitter evolution in scalar assisted $f(R)-\phi$ gravity {\it Int. J. Geom. Meth. Mod. Phys.} \textbf{15} (2018), No. 12, 1850212    (arXiv:1808.04674 [gr-qc]).
\bibitem{Odintsov2018b} S. D. Odintsov and V. K. Oikonomou, Study of finite-time singularities of loop quantum cosmology interacting multifluids, {\it Phys. Rev. D} \textbf{97} (2018), 124042 (arXiv:1806.01588 [gr-qc])
\bibitem{Bahamonde2018} S. Bahamonde, C. G. Boehmer, S. Carloni, E. J. Copeland, W. Fang and N. Tamanini, Dynamical systems applied to cosmology: Dark energy and modified gravity, {\it Phys. Rept.} \textbf{775-777} (2018), 1-122  (arXiv:1712.03107 [gr-qc]).


\bibitem{Langlois2} D. Langlois, Brane cosmology: An Introduction, {\it Prog. Theor. Phys. Suppl.} \textbf{148} (2003), 181-212.
\bibitem{Brax1} P. Brax and C. van de Bruck, Cosmology and brane worlds: A Review, {\it Class. Quant. Grav.} \textbf{20} (2003), R201-R232.
\bibitem{Langlois3} D. Langlois, Cosmology of brane - worlds, arXiv: astro-ph/\textbf{0403579}.
\bibitem{Maartens1} R. Maartens, Brane world gravity, {\it Living Rev. Rel.} \textbf{7} (2004), 7.
\bibitem{Bowcock1} P. Bowcock, C. Charmousis and R. Gregory, General brane cosmologies and their global space-time structure, {\it Class. Quant. Grav.} \textbf{17} (2000), 4745-4764.


\bibitem{Kaeonikhom1} C. Kaeonikhom, D. Singleton, S. V. Sushkov, and N. Yongram,  Dynamics of Dirac-Born-Infeld dark energy interacting with dark matter, {\it Phys. Rev. D} \textbf{86} (2012), 124049 (arXiv:1209.5219 [gr-qc]).



\bibitem{Sahni1} V. Sahni, and L.-M. Wang, A New cosmological model of quintessence and dark matter, {\it Phys. Rev. D} \textbf{62} (2000),  103517  (astro-ph/9910097).
\bibitem{Sahni2} V. Sahni and A. Starobinsky, The Case for a positive cosmological Lambda term, {\it Int. J. Mod. Phys. D} \textbf{9} (2000), 373-444;
                 L. A. Urena-Lopez and T. Matos, A New cosmological tracker solution for quintessence, {\it Phys. Rev. D} \textbf{62} (2000), 081302;
                 S. A. Pavluchenko, Generality of inflation in closed cosmological models with some quintessence potentials, {\it Phys. Rev. D} \textbf{67} (2003), 103518.

\bibitem{Salcedo1} R. Garcia-Salcedo, T. Gonzalez, F. A. Horta-Rangel, I. Quiros, and D. Sanchez-Guzman, Introduction to the application of dynamical systems theory in the study of the dynamics of cosmological models of dark energy,  {\it Eur. J. Phys.} \textbf{36} (2015), no.2, 025008 (arXiv:1501.04851 [gr-qc]).

\bibitem{Demianski} M. Demianski, E. Piedipalumbo, D. Sawant and L. Amati, Cosmology with gamma-ray bursts: II Cosmography challenges and cosmological scenarios for the accelerated Universe, {\it Astron. Astrophys.} \textbf{598} (2017), A113 (arXiv:1609.09631 [astro-ph.CO]).
















%%%%%%%%%%%%%%%%%%%%%%%%%%%%%%%%%%%%%%%%%%%%%%%%%%%%%%%%%%%%%%%%%%%%%%%%%%%%%%%%%%%%%%%%%%%%%%%%%%%%%%%%%%%%%%%%%

%%%%%%%%%%%%%%%%%%%%%%%%%%%%%%%%%%%%%%%%%%%%%%%%%%%%%%%%%%%%%%%%%%%%%%%%%%%%%%%%%%%%%%%%%%%%%%%%%%%%%%%%%%%%%%%%%%%

\end{thebibliography}
\end{document}